\documentclass[english,usenatbib]{mn2e}
\usepackage[a4paper]{geometry}
\usepackage{psfig}
\usepackage[dvips]{graphicx}
\usepackage{hyperref}

\usepackage{amsmath}

\newcommand{\LCDM}{$\Lambda$CDM}
\newcommand{\msun}{\mbox{${\rm M}_{\odot}$}}
\newcommand{\hmsun}{\mbox{${\rm h^{-1}\, M}_{\odot}$}}

\def\lesssim{\lower.5ex\hbox{$\; \buildrel < \over \sim \;$}}
\def\gtrsim{\lower.5ex\hbox{$\; \buildrel > \over \sim \;$}}

\title[Relating Galaxy Size to Halo Size]{The Relationship between Galaxy and Dark Matter Halo Size from $z\sim 3$ to the present}

\author[R. S. Somerville et al.] {
Rachel S. Somerville$^{1,2}$, Peter Behroozi$^{3}$, Viraj Pandya$^{4}$,
Avishai Dekel$^{5}$, 
\newauthor S. M. Faber$^{4}$, Adriano Fontana$^{6}$,
Anton M. Koekemoer$^{7}$,
\newauthor David C. Koo$^{4}$,  P.~G. P\'erez-Gonz\'alez$^{8}$, Joel R. Primack$^{9}$, Paola Santini$^{6}$,
\newauthor Edward N. Taylor$^{10}$,
Arjen van der Wel$^{11}$\\
$^1$Department of Physics and Astronomy, Rutgers, The State University of New Jersey, 136 Frelinghuysen Rd, \\Piscataway, NJ 08854, USA; 
\href{mailto:somerville@physics.rutgers.edu}{somerville@physics.rutgers.edu}\\
$^2$Center for Computational Astrophysics, Flatiron Institute, 162 5th Ave, New York, NY 10010, USA\\
$^3$Hubble Fellow, Department of Astronomy, University of California, Berkeley, CA 94720, USA\\
$^4$UCO/Lick Observatory, Department of Astronomy and Astrophysics, University of California, Santa Cruz, CA 95064, USA \\
$^{5}$Center for Astrophysical and Planetary Science, Racah Institute of Physics, The Hebrew University, Jerusalem 91904, Israel\\
$^{6}$ INAF - Osservatorio Astronomico di Roma, via di Frascati 33,
00078 Monte Porzio Catone, Italy\\
$^{7}$Space Telescope Science Institute, 3700 San Martin Drive, Baltimore, MD 21218, USA\\
$^{8}$Departamento de Astrof\'isica, Facultad de CC. F\'isicas, Universidad Complutense de Madrid, E-28040 Madrid, Spain\\
$^{9}$Department of Physics, University of California, Santa Cruz, CA 95064, USA \\
$^{10}$Centre for Astrophysics and Supercomputing, Swinburne University of Technology, PO Box 218, Hawthorn 3122, Australia\\
$^{11}$Max Planck Institute for Astronomy, Konigstuhl 17, D-69117 Heidelberg, Germany\\
} 
\begin{document}

\maketitle

\begin{abstract}
We explore empirical constraints on the statistical relationship between the
radial size of galaxies and the radius of their host dark matter halos
from $z\sim 0.1$--3 using the GAMA and CANDELS surveys. We map dark
matter halo mass to galaxy stellar mass using relationships
from abundance matching, applied to the Bolshoi-Planck dissipationless
N-body simulation. We define SRHR$\equiv r_e/R_h$ as the ratio of
galaxy radius to halo virial radius, and SRHR$\lambda \equiv
r_e/(\lambda R_h)$ as the ratio of galaxy radius to halo spin
parameter times halo radius. At $z\sim 0.1$, we find an average value
of SRHR $\simeq 0.018$ and SRHR$\lambda \simeq 0.5$ with very little
dependence on stellar mass.  SRHR and SRHR$\lambda$ have a weak
dependence on cosmic time since $z\sim 3$. SRHR shows a mild decrease
over cosmic time for low mass galaxies, but increases slightly or does
not evolve for more massive galaxies.  We find hints that at high
redshift ($z\sim 2$--3), SRHR$\lambda$ is lower for more massive
galaxies, while it shows no significant dependence on stellar mass at
$z\lesssim 0.5$. We find that for both the GAMA and CANDELS samples,
at all redshifts from $z\sim 0.1$--3, the observed conditional size
distribution in stellar mass bins is remarkably similar to the
conditional distribution of $\lambda R_h$.  We discuss the physical
interpretation and implications of these results.
\end{abstract}

\begin{keywords}
galaxies: evolution - galaxies: formation - galaxies: structure - galaxies: high redshift
\end{keywords}

\section{Introduction}
\label{sec:intro}

Our standard modern paradigm of galaxy formation posits that galaxies
form within dark matter halos, and much recent work has focussed on
empirically relating the observable properties of galaxies with those
of their host halos. While there are many ways to approach this
problem, a commonly used approach to constrain the relationship
between the stellar mass (or luminosity) of galaxies and the mass of
their host dark matter halos (the SMHM relation) is (sub-)halo
abundance matching
\citep[SHAM;][]{conroy:2006,guo:2010,behroozi:2010,moster:2010,behroozi:2013,moster:2013}.
The ansatz of such models is that a galaxy global property such as
stellar mass is tightly correlated with the host halo mass (or other
property, such as internal velocity). One can then ask: what sort of
mapping between galaxy property ($m_*$) and halo property ($M_h$)
would allow us to match the predicted abundance (from
\LCDM\ cosmological simulations) of halos with mass $M_h$ with the
observed abundance of galaxies with mass $m_*$, at any given redshift
$z$? The abundance matching formalism has proven to be extremely
powerful, and agrees well with other constraints from clustering,
satellite kinematics, and gravitational lensing \citep[see
  e.g.][]{behroozi:2013}.

Observationally, it is well known that galaxy stellar mass and
luminosity are strongly correlated with structural properties such as
radial size
\citep{kormendy:1977,courteau:2007,shen:2003,bernardi:2010,lange:2015},
although there is a significant dispersion in radial size at a given
stellar mass or luminosity \citep{dejong:2000,shen:2003}. There have
been a great many studies of the cosmic evolution of the galaxy
size-mass (and size-luminosity) relation
\citep{lilly:1998,simard:1999,giavalisco:1996,lowenthal:1997,ravindranath:2004,ferguson:2004,barden:2005,trujillo:2006,vandokkum:08}. With
the installation of Wide Field Camera 3 on the Hubble Space Telescope,
and the completion of extensive multi-wavelength surveys such as
CANDELS \citep{grogin:2011,koekemoer:2011} and 3D-HST
\citep{skelton:2014,momcheva:2016}, we have gained the ability to
study galaxy structure at high redshift with unprecedented fidelity
and robustness. These large and highly complete surveys have allowed
us to study the dependence of the size-mass relation and its cosmic
evolution on galaxy properties such as morphology or star formation
activity
\citep{newman:2010,damjanov:2011,cassata:2011,vanderwel:2014,vandokkum:2015}.

Considerable effort has been devoted to attempting to understand the
physical origin of the size-mass relation and its evolution. Both dark
matter and diffuse gas acquire angular momentum via tidal torques and
mergers
\citep{peebles:1969,White:1984,porciani:2002,vitvitska:2002}. The
specific angular momentum is often written using the dimensionless
spin parameter\footnote{An alternative definition due to
  \citet{bullock:2001b} is $\lambda_B \equiv J(\sqrt{2}
  MVR)^{-1}$. Unless otherwise specified, $\lambda$ denotes the
  Peebles definition in this work, but we also compare with the
  Bullock definition denoted by $\lambda_B$.}:
\begin{equation}
  \lambda = \frac{J |E|^{1/2}}{GM^{5/2}}
  \label{eqn:spin}
\end{equation}
where $J$ is the total angular momentum, $E$ is the total energy, $G$
is Newton's gravitational constant and $M$ is the total mass
\citep{peebles:1969}.  In the classical picture, diffuse gas acquires
about the same amount of specific angular momentum as the dark matter,
and conserves most of this angular momentum as it cools, collapses,
and forms stars. The very simplest, most na\"{i}ve model of disk
formation makes the following assumptions: halos are spherical and
have a singular isothermal density profile $\rho \propto r^{-2}$ with
all particles on circular orbits; gas collapses to form a disk with an
exponential radial profile, conserving its angular momentum;
self-gravity is neglected. Under this set of assumptions, we expect
the disk's exponential scale radius to be given by:
\begin{equation}
  r_d = \frac{1}{\sqrt{2}} \lambda R_h
  \label{eqn:rd_iso}
\end{equation}
where $R_h$ is the virial radius of the dark matter halo
\citep[e.g.][]{mo:1998}. Numerous refinements to this simplest model
have been presented in the literature. These include the deviation of
dark matter halo profiles from isothermal spheres, modification of the
inner halo profiles by self-gravity or energy input by stars or an
active black hole, and transfer of angular momentum during the disk
formation process or due to mergers
\citep[e.g.][]{blumenthal:1986,mo:1998,dutton:2007,somerville:2008a,shankar:2013,porter:2014}. These
more detailed models are discussed further in
Section~\ref{sec:physics}, but we note here that the expression
derived under presumably more realistic assumptions retains the
proportionality $r_d \propto \lambda R_h$ (see Eqn.~\ref{eqn:mmw}).

Most semi-analytic models (SAM) of galaxy formation
  \citep[e.g.][]{kauffmann:1996,SP:1999,cole:2000,Croton:2006,monaco:2007,somerville:2008b,Benson:2012,somerville:2015,henriques:2015,croton:2016,lacey:2016}
  adopt this ``angular momentum partition'' ansatz, and use an
  expression like Eqn.~\ref{eqn:rd_iso} or variants such as those
  discussed in Section~\ref{sec:physics}, to model the sizes of
  galactic disks. Not all such models that adopt this ansatz have
  explicitly published their predicted size-mass relations, but some
  models have shown reasonable success in reproducing the observed
  size-mass relation for the stars in disks over the redshift range
  $z\sim 0$--2
  \citep{somerville:2008a,dutton:2011a,dutton:2012}. \citet{popping:2014}
  compared their SAM predictions with the sizes of cold gas disks of
  molecular hydrogen (traced by CO) or neutral atomic hydrogen (HI),
  finding good agreement at $z\sim 0$--2.
Some SAMs also include a model for the sizes of spheroids formed in
mergers and disk instabilities
\citep{shankar:2010,shankar:2013,porter:2014}. As spheroids form out
of disks in these models, the sizes of the spheroids depend on the
sizes of their disky progenitors. Models that include the effects of
dissipation have been shown to be successful at reproducing the slope
and normalization of the size-mass relation for spheroid-dominated
galaxies and its evolution from $z\sim 2$--0, while models that do not
account for the effects of dissipation do not fare so well
\citep{shankar:2010,shankar:2013,porter:2014}. 

There has also been extensive study of the radial sizes of galaxies
(particularly disks) predicted by numerical cosmological hydrodynamic
simulations. Indeed, correctly reproducing the galaxy size-mass
relation and its evolution poses a stringent challenge for numerical
simulations. Early simulations were plagued by an ``angular momentum
catastrophe'', in which galaxies were much too compact for their mass
\citep{navarro_steinmetz:2000,sommer_larson:1999,steinmetz_navarro:2002}. In
these simulations, stellar disks ended up with a much smaller angular
momentum than that of their dark matter halo due to large angular
momentum losses during the formation process.

More recently, improvements in hydrodynamic solvers, numerical
resolution, and sub-grid treatments of star formation and stellar
feedback have enabled at least some hydrodynamic simulations to
reproduce the observed size-mass relation for disks in ``zoom-in''
simulations
\citep{governato:2004,governato:2007,christensen:2012,guedes:2011,aumer:2014},
and its evolution since $z\sim 1$ \citep{brooks:2011}. However, the
predicted sizes of galaxies in numerical simulations are very
sensitive to the details of the sub-resolution prescriptions for star
formation and feedback processes --- different implementations of
feedback that all reproduce global galaxy properties (such as stellar
mass functions) can produce galaxies with very different size-mass
relations and morphologies
\citep{scannapieco:2012,uebler:2014,schaye:2015,crain:2015,genel:2015,agertz:2016}. For
example, the EAGLE simulations \citep{schaye:2015}, which were tuned
to reproduce the size-mass relation for disks at $z\sim0$, appear to
be consistent with observational measurements of the size-mass
relation for both star forming and quiescent galaxies back to $z\sim
2$ \citep{furlong:2015}. However, the Illustris simulations
\citep{vogelsberger:2014}, which did not use radius as a tuning
criterion, produce galaxies that are about a factor of two larger than
observed galaxies at a fixed stellar mass
\citep{snyder:2015,furlong:2015}. Several studies have shown that
  various assumptions of the angular momentum partition plus adiabatic
  contraction type models \citep[e.g.][]{mo:1998} are violated in
  numerical hydrodynamic simulations
  \citep{sales:2009,stevens:2017,desmond:2016}. \citet{desmond:2016}
  showed that in the EAGLE simulations, galaxy size is almost
  uncorrelated with halo spin.

Clearly the validity of the classical angular momentum partition
ansatz contained in Eqn.~\ref{eqn:rd_iso} --- that galaxy size is
strongly correlated with the spin and radius of the host dark matter
halo --- lies at the heart of this issue.
There are many reasons to expect that there would \emph{not} be a
simple one-to-one correspondence between the spin of the cold baryons
(stars and cold gas in the interstellar medium) in galaxies
$\lambda_{\rm galaxy}$ and the spin of the dark matter halo within the
virial radius $\lambda_h$. These can be grouped into two categories:
1) the angular momentum of the baryons that end up in the galaxy may
not be an unbiased sample of the initial angular momentum of the halo;
and 2) angular momentum may be lost or gained by the baryonic
component during the formation process. With regard to 1), numerical
cosmological simulations have shown that most of the cold gas that
forms the fuel for stars that end up in disks, in particular, is not
accreted from a spherical hot halo in virial equilibrium, but rather
along cold filaments \citep{brooks:2009}.  This gas has 2-5 times more
specific angular momentum than the dark matter halo when it is first
accreted into the galaxy
\citep{stewart:2013,danovich:2015}. Furthermore, after gas has
accreted into the disk, a large fraction of it is ejected again by
stellar-driven winds. Low-angular momentum material is
preferentially removed, and ejected gas can be torqued up by
gravitational fountain effects \citep{brook:2012,uebler:2014}. With
regard to 2), angular momentum may be transferred from the baryonic
component to the dark matter halo by mergers
\citep{mihos:1995,dekel:2006,covington:2008,hopkins:09a} or internal
processes such as viscosity and disk instabilities
\citep{dekel:2009,Dekel_Burkert:2014,danovich:2015}. \citet{zjupa:2017}
find an overall enhancement of the spin of baryons in galaxies
relative to halo spin of a factor of 1.8 at $z=0$ in the Illustris
simulations.

A galaxy-by-galaxy comparison of the spin of either the stars or cold
baryons in galaxies with the spin of their host halo shows a very
rough correlation between $\lambda_{\rm galaxy}$ and $\lambda_h$, but
with a scatter of about two orders of magnitude
\citep{zavala:2016,teklu:2015}. This correlation is found to depend on
galaxy morphology, with $\lambda_{\rm galaxy}$ lying systematically
below $\lambda_h$ for spheroid dominated galaxies, and disk galaxies
lying around the $\lambda_{\rm galaxy} \simeq \lambda_h$ line
\citep{teklu:2015}. Similarly, \citet{zjupa:2017} find a halo mass
dependence for the ratio of baryonic to halo spin. \citet{dekel:2013}
compute the ratio of the disk radius to the halo radius $r_{d}/R_h$
for 27 cosmological zoom-in simulations of moderately massive halos,
finding a mean value of SRHR$=0.06$ at $z\sim 4$, declining to 0.05 at
$z\sim 2$ and 0.04 at $z\sim 1$. The 68th percentile halo-to-halo
dispersion around these values is large, around 50\%. 

Given that the relationship between halo and galaxy angular momentum
is so sensitive to the still very uncertain details of sub-grid
feedback recipes in numerical simulations, it appears useful to
investigate purely empirical constraints on this relationship and its
dependence on galaxy or halo mass and cosmic time. In this paper, we
investigate the statistical relationship between the observed size
(stellar half-light or half-mass radius) of galaxies and the inferred
size (virial radius) of their dark matter halos via stellar mass
abundance matching. In analogy to the stellar-mass-halo-mass relation
(SMHM), we term this the stellar-radius-halo-radius (SRHR)
relation. We define the quantity SRHR $\equiv r_e/R_h$, where $r_e$ is
the half-mass or half-light radius of the galaxy and $R_h$ is the
virial radius of the halo\footnote{Both $r_e$ and $R_h$ need to be
  defined more carefully. We discuss this in later sections.}. As we
wish to explore the relationship between the \emph{angular momentum}
of the halo and that of the galaxy, we further define and investigate
the quantity SRHR$\lambda \equiv r_e/(\lambda R_h)$.  As discussed
above, we expect that SRHR and SRHR$\lambda$ will vary from galaxy to
galaxy. With our approach, we can primarily constrain the median or
average value of these parameters in bins of stellar mass and
redshift. We also make an attempt to constrain the galaxy-to-galaxy
dispersion of these quantities, but this is more indirect.

To achieve this goal, we use the SHAM approach to assign stellar
masses to dark matter halos from the Bolshoi-Planck dissipationless
N-body simulation \citep{rodriguez-puebla:2016}. Using the observed
relationship between stellar mass and radius derived from
observations, we then infer the median or average value of SRHR and
SRHR$\lambda$ in stellar mass bins. In addition, we can use the
conditional size distribution (the distribution of galaxy radii in
stellar mass bins) to place limits on the allowed amount of
galaxy-to-galaxy scatter in SRHR$\lambda$. We apply this approach to a
sample of nearby galaxies ($z\sim 0.1$) taken from the GAMA survey
\citep{driver:2011,liske:2015}, and also to observational measurements
of galaxy radii from the CANDELS survey \citep[][hereafter
  vdW14]{vanderwel:2014} over the redshift range $0.1<z<3$. Although
there have been several observational studies of galaxy size evolution
at higher redshifts, up to $z\sim 8$
\citep[e.g.][]{huang:2013,shibuya:2015,oesch:2010,curtis-lake:2016},
we do not attempt to extend the current study to redshifts greater
than three, for several reasons: we do not have reliable stellar mass
estimates, available light profiles probe the rest-UV, which may not
accurately reflect the radial distribution of stellar mass, and
selection and measurement effects may have a larger impact at these
redshifts \citep{curtis-lake:2016}.

\citet[][K13]{kravtsov:2013} used a similar approach based on
abundance matching to relate stellar mass to halo mass, and then
demonstrated the surprising result that the observed sizes (half-light
radii) of nearby galaxies were consistent with being on average
\emph{linearly proportional} to their halo virial radii. Still more
surprisingly, he found that the linear proportionality held, with the
same scaling factor, over many orders of magnitude in mass and size,
and for galaxies of diverse morphology, from dwarf spheroidals and
irregulars to spirals to giant ellipticals.

Recent work by \citet{shibuya:2015} has similarly examined the
relationship between galaxy size and halo size out to high redshift
using abundance matching.  Another recent work by
\citet[][H17]{Huang:2017} carries out a related study of galaxy size
versus halo size using the same CANDELS dataset used here. Our study
is complementary to these previous works in several respects, and we
discuss these differences in detail, including presenting a direct
comparison with the analysis of H17, in Section~\ref{sec:discussion}.

Some of the more important new aspects of our work are as follows. In
this paper, we ``forward model'', taking halos from a cosmological
N-body simulation to the observational plane, while many other studies
(e.g. K13, \citet{shibuya:2015}, and H17) ``backwards model'', taking
the observed galaxies to theory space using abundance matching or by
inverting a SMHM relation. As we will show explicitly, backwards
  modeling can suffer from substantial biases in the presence of
  dispersion in the SMHM relation, while our approach explicitly
  accounts for that dispersion.
  In addition, we carry out our analysis on a local galaxy sample from
  GAMA and the high redshift CANDELS observations in a consistent
  manner, while this has not been done in previous studies. Thirdly,
  we carry out a detailed comparison with the conditional size
  \emph{distributions} in stellar mass bins, rather than just the mean
  or median size.

The structure of the remainder of this paper is as follows. In
Section~\ref{sec:model}, we describe the simulations and sub-halo
abundance matching model (SHAM) that we use. We summarize the
observational datasets that we make use of in
Section~\ref{sec:observations}.  We present our main results in
Section~\ref{sec:results}. We discuss the interpretation of our
results, and compare our results with those of other studies in the
literature, in Section~\ref{sec:discussion}. We summarize and conclude
in Section~\ref{sec:summary}. Two Appendices present supplementary
results on observational size-mass relations and definitions of halo
structural parameters.

\section{Simulations and sub-halo abundance matching model}
\label{sec:model}

\subsection{Simulations and Dark Matter Halos}
We use the redshift $z=0.10$ snapshot from the Bolshoi-Planck
simulation \citep{rodriguez-puebla:2016} and CANDELS mock lightcones
extracted from the Bolshoi-Planck simulation (Somerville et al. in
prep). Bolshoi-Planck contains $2048^3$ particles within a box that is
250 h$^{-1}$ comoving Mpc on a side, and has a particle mass of
$1.5\times 10^8\, \hmsun$. The Plummer equivalent gravitational
softening length is 1 h$^{-1}$ kpc. The Bolshoi-Planck simulations
adopt the following values for the cosmological parameters:
$\Omega_{m,0}=0.307$, $\Omega_{\Lambda,0} = 0.693$,
$\Omega_{b,0}=0.048$, $h=0.678$, $\sigma_8=0.823$, and
$n_s=0.96$. These are consistent with the Planck 2013 and 2014
constraints \citep{planck:2014}. We adopt these values throughout our
analysis.

Dark matter halos and sub-halos have been identified using the
ROCKSTAR algorithm \citep{rockstar}; many properties of the halos in
Bolshoi-Planck are presented in \citet{rodriguez-puebla:2016}.
We make use of the halo virial mass, halo virial radius, and halo spin
parameter. The Bolshoi-Planck halo catalogs provide both the
``Peebles'' and ``Bullock'' definition of the spin parameter; we use
the Peebles definition (Eqn.~\ref{eqn:spin}) unless specified
otherwise but also compare the results of using the Bullock
  definition. Following \citet{behroozi:2013}, we define halo virial
mass and radius within spherical overdensity $\Delta_{\rm vir}$ times
the critical density, where $\Delta_{\rm vir}$ is given by the fitting
function presented in Eqn.~6 of \citet{bryan:1998}:
\begin{equation}
\Delta_{\rm vir} = 18\pi^2 + 82x - 39x^2
\end{equation}
where $x=\Omega(z)-1$, with $\Omega(z)$ the matter density relative to
the critical density at redshift $z$. Note that
\citet{rodriguez-puebla:2016} write $\Delta_{\rm vir} (z) = (18\pi^2 +
82x - 39x^2)/\Omega(z)$ and then write $M_{\rm vir} = \frac{4\pi}{3}
\Delta_{\rm vir} \rho_m R_{\rm vir}^3$, making it appear that halo
mass is defined relative to the average matter density $\rho_m$ rather
than the critical density $\rho_c(z)$. However, since $\rho_m(z) =
\Omega(z) \rho_c(z)$, quick inspection reveals that their expression
is equivalent to the \emph{original} Bryan \& Norman expression with
$M_{\rm vir} = \frac{4\pi}{3} \Delta_{\rm vir} \rho_{\rm crit} R_{\rm
  vir}^3$, and indeed equivalent to our definition.  See
Section~\ref{sec:discussion} for further discussion of the
implications of different halo definitions. We emphasize that the
  measurement of the halo spin parameter, halo mass and radius, and
  the sub-halo abundance matching model used in this work all adopt a
  consistent set of cosmological parameters and halo definitions.

\subsection{Relating (sub-)Halos to Galaxies}

\begin{figure*} 
\begin{center}
\includegraphics[width=\textwidth]{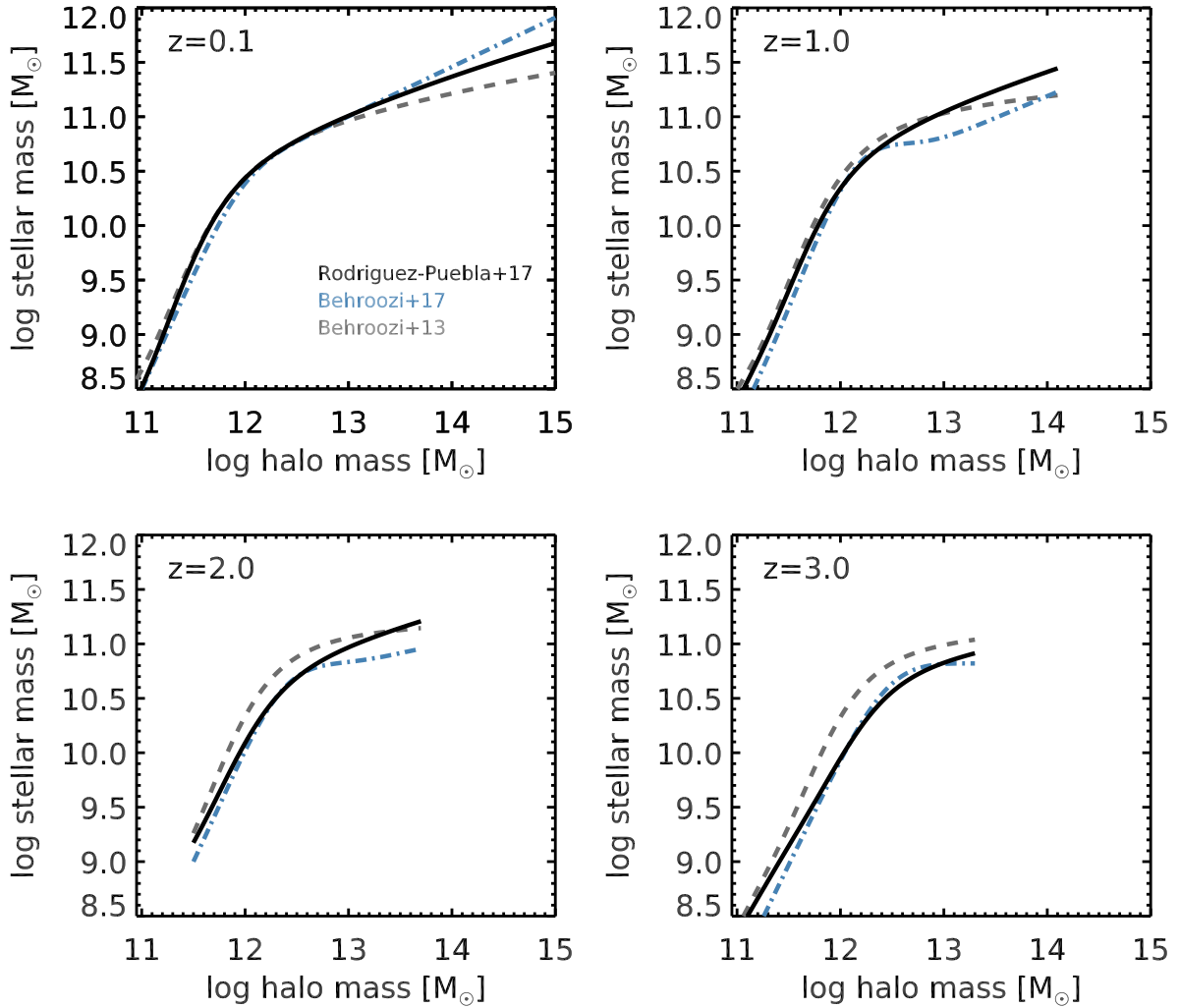}
\end{center}
\caption{Stellar mass versus halo mass relation derived from abundance
  matching, shown at several different redshifts as indicated in the
  panels. The black solid line shows the relation from \citet{rp17},
  which is the fiducial relation used in this work. The blue
  dot-dashed line and gray dashed line show the relations
  derived by Behroozi et al. (in prep) and \citet{behroozi:2013},
  respectively. 
  \label{fig:smhm}}
\end{figure*}

We make use of results from the well-established technique of sub-halo
abundance matching (SHAM) to assign stellar masses to each halo and
sub-halo in the Bolshoi-Planck catalogs. We show several recent SMHM
relations derived from sub-halo abundance matching at several relevant
redshifts in Fig.~\ref{fig:smhm}. This figure illustrates the
differences in the SMHM derived by two different sets of authors
(\citet{rp17} and Behroozi et al. in prep, hereafter RP17 and B17
respectively) based on the same underlying (sub)-halo
distributions. Differences can arise from the choice of observations
used to constrain the SHAM, as well as details of the
methodology. Note that the analysis of \citet{behroozi:2013} was based
on the original Bolshoi simulations, rather than Bolshoi-Planck, and
adopts slightly different cosmological parameters than those used by
RP17 and B17. The difference in cosmological parameters explains why
the \citet{behroozi:2013} SMHM relation is slightly higher at
$z\gtrsim 1$ than those derived by RP17 and B17. The difference
between the RP17 and B17 SMHM at large halo masses mainly arises from
the use of different observational determinations of the stellar mass
function. We can see from Fig.~\ref{fig:smhm} that for the
  stellar mass range on which we focus in our study ($m_* \gtrsim 10^9
  \, \msun$), galaxies are hosted by halos that are well resolved in the
  Bolshoi-Planck simulations, with at least several thousand particles
  within the virial radius.

The SMHM relation has dispersion both due to intrinsic scatter in the
relation, and due to observational stellar mass errors.  Following
RP17, we adopt an intrinsic scatter of $\sigma_h=0.15$ dex and a
scatter due to observational stellar mass measurement errors of
$\sigma_*=0.1 + 0.05z$, where $z$ is redshift. These choices are
consistent with the constraints summarized by
\citet{tinker:2017}. Operationally, we assign $\langle \log
m_*(M_h)\rangle$ from Eqn. 25-33 of RP17, then add to this number a
Gaussian random deviate with standard deviation $\sigma_T =
(\sigma_h^2 + \sigma_*^2)^{1/2}$. For halo mass, we use the maximum mass
along the halo's history, as in RP17. For distinct (non-sub) halos,
this is generally equivalent to the standard virial mass, while for
sub-halos, this has been shown to produce better agreement with
clustering measurements. We assume that satellite galaxies obey the
same SMHM relationship as central galaxies. While this may not be
precisely correct \citep{zheng:2005}, the great majority of galaxies
in the mass range we study are central galaxies, so our results should
not be very sensitive to this assumption.

Consistency with observed clustering measurements is an important
  check of SHAMs.  A similar SHAM has been shown to be consistent with
  clustering measurements by \citet{rodriguez-puebla:2015}, and the
  updated RP17 SHAM is also consistent with observed galaxy two-point
  correlation functions from $z\sim 0.1$--1 (A. Rodriguez-Puebla,
  priv. comm.).

With stellar masses assigned to each of our halos, we can then
  use a simple but robust approach to constrain the median
  relationship between galaxy size and halo size. We bin our SHAM
  sample in stellar mass and compute the medians of the halo radius
  $\langle R_h(m_*)\rangle$ and spin times halo radius $\langle
  \lambda R_h(m_*)\rangle$. Similarly, we compute the median observed
  galaxy size in the same stellar mass bins, $\langle r_e(m_*) \rangle$. We
  then obtain SRHR$=\langle r_e \rangle/ \langle R_h\rangle$
  and SRHR$\lambda = \langle r_e \rangle/ \langle \lambda
  R_h\rangle$. We use medians as our default, but also repeat our
  analysis using means, finding qualitatively similar conclusions. 

We note that a distinction is frequently made between spheroid-
  and disk-dominated galaxies, or star forming and quiescent galaxies,
  in discussing their sizes. Some previous studies (e.g. K13, H17)
  present results for the relationship of galaxy size to halo size for
  samples of different galaxy types, using a common abundance matching
  relation. We make a deliberate choice \emph{not} to divide galaxies
  by type in our study, as there is strong evidence that star
  forming/disk dominated galaxies and quiescent/red galaxies have
  significantly different SMHM relations
  \citep[e.g.][]{rodriguez-puebla:2015}. Moreover, it is possible that
  disk and spheroid dominated galaxies may arise from halos with
  different spin parameter distributions. In order to avoid making
  assumptions about possible differences in the properties of
  halos that host different types of galaxies, we simply compute our
  results in stellar mass bins.

Another interesting issue is whether the size-mass relation for
galaxies depends on the larger scale environment (e.g., on scales
larger than the halo virial radius). Similarly, looking into this
issue requires knowledge of whether the SMHM relation is universal or depends
on environment. We intend to investigate this is future works.

\section{Observational Data}
\label{sec:observations}

\subsection{GAMA}
\label{sec:observations:gama}

To characterize nearby galaxies, we make use of the catalogs from Data
Release 2 (DR2) of the Galaxy And Mass Assembly survey
\citep[GAMA;][]{liske:2015}, covering 144 square degrees. GAMA is an
optically selected, multi-wavelength survey with high spectroscopic
completeness to $r<19.8$ mag (two magnitudes deeper than the Sloan
Digital Sky Survey; SDSS). We make use of stellar mass estimates from
\citet{taylor:2011}, and structural properties (semi-major axis
half-light radius and S{\'e}rsic parameter) from the analysis of
\citet{kelvin:2012} using the GALFIT code \citep{galfit}.  We restrict
our sample to a redshift range $0.01<z<0.12$ and require $r$-band
GALFIT quality flag = 0 (good fits only). In addition, we discard
galaxies with S{\'e}rsic index $n_s< 0.3$ or $n>10$, as these are
typically signs of unreliable fits, and we also exclude galaxies with
sizes $r_e < 0.5$ FWHM. The FWHM is set by the seeing of SDSS, for
which we adopt an average value of 1.5''. We adopt the results of
structural fits in the observed $r$-band (due to the relatively small
redshift range probed by GAMA, k-corrections should not be
needed). After these cuts, we have 13,771 galaxies in our GAMA sample.

\subsection{CANDELS}
\label{sec:observations:candels}

CANDELS is anchored on HST/WFC3 observations of five widely-spaced
fields with a combined area of about 0.22 sq. deg. An overview of the
survey is given in \citet{grogin:2011} and \citet{koekemoer:2011}. The
data reduction and cataloging for each of the fields is presented in
\citet[][COSMOS]{nayyeri:2016}, \citet[][EGS]{stefanon:2017}, Barro et
al. (in preparation; GOODS-N), \citet[][GOODS-S]{guo:2013}, and
\citet[][UDS]{galametz:2013}. The primary CANDELS catalogs are
selected in F160W ($H$-band), and CANDELS has a rich ancillary
multi-wavelength dataset extending from the radio to the X-ray (see
\citet{grogin:2011} for a summary). Photometric redshifts have been
derived as described in \citet{dahlen:2013}. We make use of the
$z_{\rm best}$ redshift from the CANDELS catalog, which selects the
best available redshift estimate from spectroscopic, 3DHST grism
based, and photometric redshifts. Stellar masses are estimated by
fitting the spectral energy distributions as described in
\citet{mobasher:2015}, with further details for each field given in
\citet{santini:2015} for GOODS-S and UDS, \citet{stefanon:2017} for
EGS, \citet{nayyeri:2016} for COSMOS and Barro et al. (in prep) for
GOODS-N. The stellar masses were derived assuming a
\citet{chabrier:2003} stellar initial mass function.

Structural parameters were derived using GALFIT as described in
\citet{vanderwel:2012}. The fits were done using a single-component
S{\'e}rsic model. The effective radius that we use is the semi-major
axis of the ellipse that contains half of the total flux of the best
fitting S{\'e}rsic model.  We select CANDELS galaxies with apparent
magnitude H$_{160} < 24.5$, PHOTFLAG$=0$ (good photometry), $0 <
z_{\rm best} \leq 3.0$, GALFIT quality flag = 0 (good fits), and
stellarity parameter CLASS\_STAR $<0.8$. We discard galaxies with
relative size errors greater than a factor 0.3. There are 49241
galaxies in the catalog with H$_{160} < 24.5$ and $0 < z_{\rm best}
\leq 3.0$. Adding the photometry and stellarity criteria brings the
number down to 45015. The GALFIT quality cut further reduces the
number to 38610, and the error clipping to 28840. Note that we have
repeated our analysis without clipping the size errors, and using a
magnitude limit of 25.5. Our results do not change significantly.

In this work, we use the sizes measured from the observed H$_{160}$
image, and apply structural k-corrections to convert to rest-frame
5000\AA\ sizes. We apply the redshift and stellar mass dependent
correction for ``late type'' galaxies given by vdW14 (their Eqn. 1) to
galaxies with S{\'e}rsic index $n_s<2.5$, and apply a constant
correction $\Delta \log R_{\rm eff}/\Delta \log \lambda_{\rm eff} =
-0.25$ to galaxies with $n_s>2.5$ (again following vdW14, for ``early
type'' galaxies). Although vdW14 use a UVJ color cut, rather than
S{\'e}rsic index, to divide early and late type galaxies, and use
sizes derived from the J$_{125}$ image rather than the H$_{160}$ one
at $z<1.5$, we have confirmed that when we follow exactly the same
procedure as vdW14, we get results that are indistinguishable for the
purposes of this paper (these alternate choices were adopted simply
for convenience).  Moreover, as we show later, our results for the
size-mass relation of galaxies from $3 \la z \la 0.2$ are in very good
agreement with the published size-mass relations from vdW14 and with
the independent analysis of H17. Basic estimates of the redshift and
color dependent stellar mass completeness limits are given in
vdW14. It is important to keep in mind that our sample may be somewhat
incomplete in the three highest redshift bins. We carry out a more
detailed assessment of the magnitude, size, color, and S{\'e}rsic
dependent completeness of the CANDELS sample in Somerville et
al. (in prep).

We convert the GAMA and CANDELS angular half-light radii to physical
kpc using the same cosmological parameters quoted above (all sizes in
this work are in physical, rather than comoving, coordinates).

\subsection{Converting from projected light to 3D stellar mass sizes}
\label{sec:observations:convert}

The radii that we obtain from the GAMA and CANDELS catalogs described
above are projected (2D) half-light radii in the rest-frame $r$ or $V$
band (approximately). We can simply relate this quantity directly to
halo properties such as virial mass and virial radius, which is useful
empirically. However, in order to gain more insight into the physical
meaning of these relationships, it is useful to attempt to convert
these sizes into 3D, stellar half-mass radii.

For simplicity, we assume that the projection from 3D to 2D and the
correction from rest-frame optical light to stellar mass can be
written as two separate terms,
\begin{equation}
  r_{\rm e, obs} = f_p \, f_{k}\, r_{\rm *, 3D}
\end{equation}
where $r_{\rm e, obs}$ is the observed (projected) effective radius of
the light in a fixed rest-frame band along the semi-major axis, and
$r_{\rm *, 3D}$ is the 3D half-mass radius of the stellar mass
distribution. The factor $f_p$ corrects for projection and $f_k$
accomplishes the structural k-correction\footnote{i.e., the conversion
  from the size in one wavelength to that in another wavelength. We
  use this term in a general sense to also refer to the conversion
  from the half-light radius to the half stellar mass radius.}.

For a face-on razor thin, transparent disk, $f_p=1$. For a spheroid,
$f_p=0.68$ for a de Vaucouleurs profile ($n_s=4$) and $f_p=0.61$ for
an exponential profile \citep[$n_s=1$; ][]{prugniel:1997}. For thick
disks or flattened spheroids, $f_p$ would be intermediate between
these values. Clearly, the dependence of $f_p$ on galaxy shape could
introduce an effective dependence of $r_{\rm e, obs}/r_{\rm *, 3D}$ on
stellar mass and/or redshift, as the mix of galaxy shapes depends on
both of these quantities \citep{vanderwel:2009a,vanderwel:2014b}.
\citet{vanderwel:2014b} showed that the fraction of elongated
(prolate) galaxies increases towards higher redshifts (up to $z\sim
2$) and lower masses, such that at $z\sim 1$, at least half of all
galaxies with stellar mass $10^9\, \msun$ are elongated. This is also
seen in numerical hydrodynamic simulations \citep{ceverino:2015}.
Dust could also affect the relationship between 3D and projected
radius. 

Regarding the structural k-corrections, \citet{dutton:2011b} quote
$f_k \sim 1.3$ for low-redshift disks in the
V-band. \citet{szomoru:2013} compute stellar mass distributions using
a single color, for a sample of a couple hundred galaxies with HST
observations as well as some nearby galaxies from SDSS, covering a
redshift range $0<z<2.5$. They find average corrections in the rest
$g$-band $\log (r_{*}/r_g) \sim -0.12$ at $z=0$, $-0.14$ at
$0.5<z<1.5$, and $-0.10$ at $1.5<z<2.5$. They did not find any strong
trends with redshift, galaxy morphology, or sSFR although again their
sample was small. They saw hints of smaller values of $\log
(r_{*}/r_g)$ for high S{\'e}rsic, quiescent galaxies.

\citet{wuyts:2012} computed stellar mass distributions using
pixel-by-pixel SED fitting with an earlier release of
CANDELS/3DHST. They show distributions of $r_e$ for rest V-band light
and for stellar mass in two redshift bins: $0.5<z<1.5$ and
$1.5<z<2.5$. They find that the distribution of half-mass radii is
shifted by 0.1 to 0.2 dex with respect to that of the half-light
radii.  There does not seem to be evidence for strong redshift
evolution, and they do not discuss any dependence on galaxy mass or
type.

\citet{lange:2015} discuss the change in size from $u$ through $K$
band for the GAMA sample. To first order one can assume that the
stellar mass effective radius is the same as the K-band light
effective radius. From $g$ to $K$ band, \citet{lange:2015} find a
decrease in radius of 16\% at $m_* \sim 10^{9} \msun$ and 13\% at $m_*
\sim 10^{10} \msun$ for `late type' galaxies. For ``early types'',
they again find 13\% at $m_* \sim 10^{10} \msun$ and 11\% at $m_* \sim
10^{11} \msun$. So again, they find weak dependence of $f_k$ on type
and stellar mass. In summary, values of $f_k$ in the literature range
from $\sim 1.12$ to 1.5. Values are perhaps slightly smaller for
quiescent/early type galaxies, but there does not seem to be evidence
for a strong trend with mass or redshift.

For purposes of this work, we adopt a rough best guess value of $(f_p
f_k)_{\rm disk} = (1*1.2) = 1.2$ for disks, and $(f_p f_k)_{\rm
  spheroid} = (0.68*1.15)=0.78$ for spheroids. To get a rough idea for
how these corrections might affect our results, we apply $(f_p
f_k)_{\rm disk}$ to estimate $r_{\rm 3D, *}$ for galaxies with
S{\'e}rsic index $n_s<2.5$, and use $(f_p f_k)_{\rm spheroid}$ for
galaxies with $n_s>2.5$. Although it is known that there is not a
perfect correspondence between galaxy \emph{shape} and S{\'e}rsic
index, this at least gives us a first approximation for how the
dependence of galaxy type on stellar mass and redshift might affect
the trends we wish to study.

For reference, we show the observed (projected light) size-mass
relations and our derived 3D stellar half mass radius relations for both
the GAMA and CANDELS samples in Fig.~\ref{fig:sizemass_gama} and
Fig.~\ref{fig:sizemass_candels}. A comparison between the
observational size-mass relations used in this work and the literature
is discussed in Appendix~\ref{sec:appendix:obs}.

Lastly, we note that for a thin exponential disk, the scale radius
$r_d$ given in Eqn. ~\ref{eqn:rd_iso} is related to the half-mass
radius via $r_{1/2} = 1.68 r_d$.

\section{Results}
\label{sec:results}

\subsection{Results from the GAMA survey at $z=0.1$}

\begin{figure} 
\begin{center}
\includegraphics[width=0.5\textwidth]{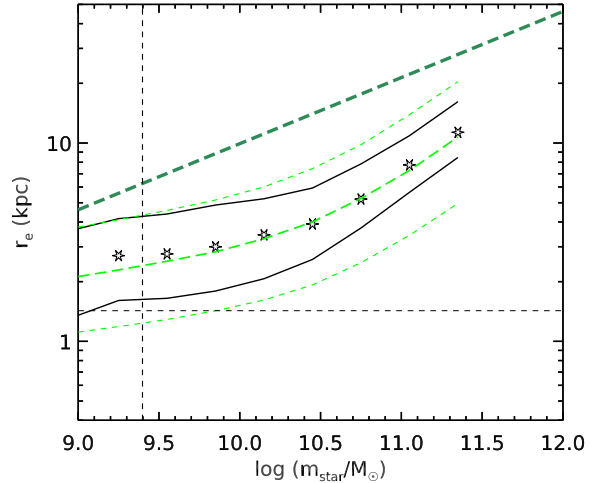}
\end{center}
\caption{The relationship between stellar mass and effective radius at
  $z\sim 0.1$.  The green dashed lines show the median and 16 and 84th
  percentiles in bins of stellar mass for the SHAM model, assuming
  $r_e = 0.5 \lambda R_h$. The black stars and lines show the median
  and 16 and 84th percentiles of the 3D half stellar mass radius for
  the GAMA $z=0.1$ sample. The horizontal dashed line shows the
  minimum size of galaxies that can be resolved at the upper redshift
  limit of the GAMA sample used here.  The dashed vertical line shows
  the 97.7\% stellar mass completeness limit for the GAMA sample.  The
  dark green dashed line shows the scaling relation for halo virial
  mass and halo virial radius (both scaled down by a factor of ten),
  illustrating that the size-mass relation for halos has a much
  steeper slope than that for galaxies.
\label{fig:sizemass_shamgama}} 
\end{figure}

Fig.~\ref{fig:sizemass_shamgama} shows the size-mass relation at
$z=0.1$ from the GAMA survey, compared with the SHAM assuming $r_e =
0.5 \lambda R_h$. Note that \citet{lange:2015} show that a stellar
mass limit of $2.5 \times 10^9 \msun$ yields a color-unbiased, 97.7\%
complete sample out to the adopted redshift limit. This limit is
indicated in Fig.~\ref{fig:sizemass_shamgama} by the vertical line.
It is somewhat remarkable how well a mass-independent value of
SRHR$\lambda$ reproduces the average size-mass relation over many
orders of magnitude in stellar mass. This result reproduces and
confirms the results already presented by K13. The median value of the
spin parameter in Bolshoi-Planck is $\langle \lambda \rangle = 0.036$,
so SRHR$\lambda=0.5$ corresponds to $r_e/R_h=0.018$ which is fairly
close to the value found by K13 ($r_e/R_h=0.015$). Also note the
steeper slope of the scaling relation between halo virial mass and
halo virial radius, compared to the observed size-mass relation for
galaxies. In this simple model, the change in slope is entirely due to
the slope of the SMHM relation.

\begin{figure*} 
\begin{center}
\includegraphics[width=\textwidth]{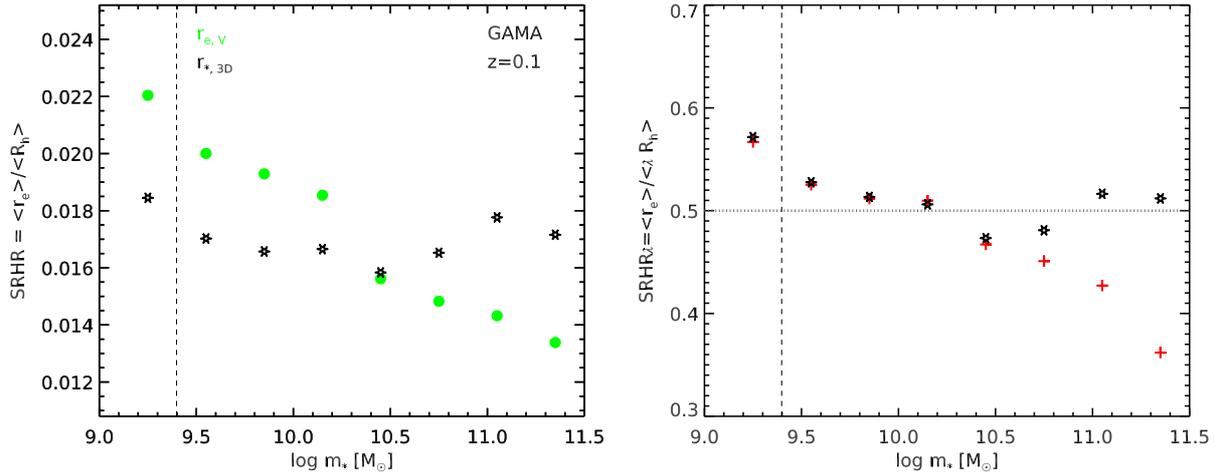}
\end{center}
\caption{Results from the GAMA survey at $z=0.1$. Left panel: Median
  galaxy radius divided by the median value of the halo virial radius
  (SRHR). Filled green circles show SRHR for the observed (projected)
  r-band half-light radius $r_e$. The dashed vertical line shows the
  97.7\% stellar mass completeness limit for the GAMA sample.  Black
  star symbols show the SRHR for the estimated 3D stellar half mass
  radius ($r_{\rm *, 3D}$).  Right panel: Median galaxy radius divided
  by the median value of the spin parameter times the halo virial
  radius (SRHR$\lambda$).  Black stars show the fiducial results,
  while red crosses show the results we would obtain if we did not
  include scatter in the SMHM relation. It is striking that the ratio
  between galaxy size and halo size remains so nearly constant over a
  wide range in stellar mass.
\label{fig:rgrh_gama}} 
\end{figure*}

In Fig.~\ref{fig:rgrh_gama} we show the ratio of the median observed
size in a stellar mass bin to the median value of $R_h$ or $\lambda
R_h$ in that same bin, again using the SHAM to link halo mass to
stellar mass. We also show the same quantity for the estimated
de-projected stellar half-mass radii. This is simply another way of
showing the results already seen above: SRHR and SRHR$\lambda$ are
nearly independent of stellar mass and have values of approximately
SRHR$=0.018$ and SRHR$\lambda=0.5$. The apparent decrease in the value
of SRHR or SRHR$\lambda$ towards larger stellar masses appears to be
mitigated by the correction from projected to 3D size.  We note that,
as found in many previous studies, we do not see any significant
dependence of the spin parameter $\lambda$ on halo mass in the
Bolshoi-Planck simulations.
 
\begin{figure*} 
\begin{center}
\includegraphics[width=\textwidth]{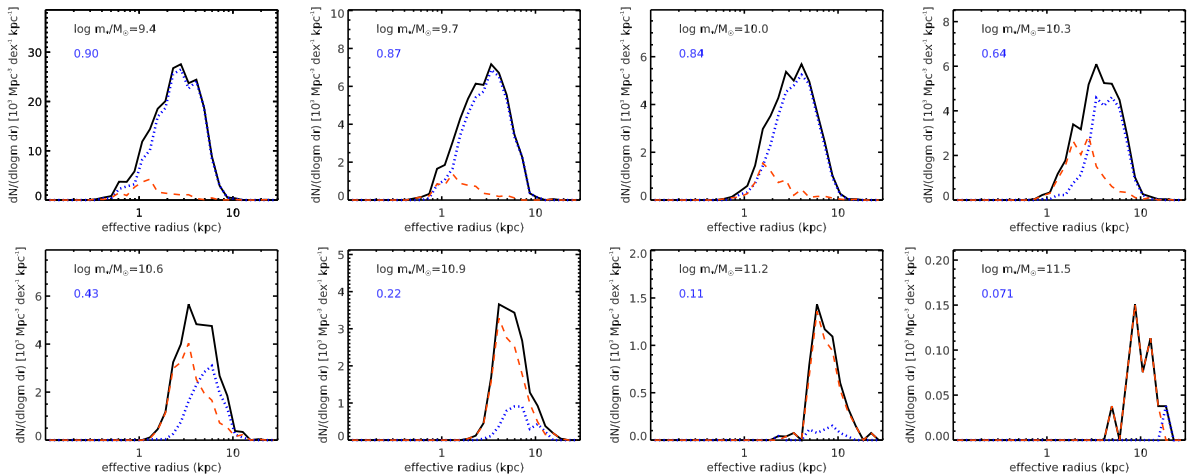}
\end{center}
\caption{Distribution functions in bins of stellar mass for the
  (2D) effective radius in the GAMA survey. The black solid lines show the
  distribution for all galaxies, the blue dotted lines show
  disk-dominated ($n_s<2.5$) galaxies, and the red dashed lines show
  spheroid-dominated ($n_s>2.5$) galaxies. The blue number in the
  upper left corner of each panel indicates the fraction of galaxies
  in that mass bin that are disk dominated ($n_s<2.5$).
\label{fig:sizedist_gama}} 
\end{figure*}

Fig.~\ref{fig:sizedist_gama} shows the conditional size distributions
(for observed half-light radii) in stellar mass bins for the GAMA
sample. We have applied a standard $V_{\rm max}$ completeness
correction, as GAMA starts to become incomplete below stellar masses
of about $10^{10} \msun$. We show the size distributions separately
for galaxies with S{\'e}rsic $n_s<2.5$, which should correspond
approximately to disk-dominated galaxies, and with S{\'e}rsic
$n_s>2.5$, which should be spheroid-dominated. We show this to
emphasize that in the lowest stellar mass bins we consider, the
distribution is dominated by $n_s<2.5$, presumably disk-dominated
galaxies, while in the highest stellar mass bins shown, the
distribution is dominated by $n_s>2.5$ spheroid dominated
galaxies. The fraction of galaxies in the bin with $n_s<2.5$ is shown
in each panel.

\begin{figure*} 
\begin{center}
\includegraphics[width=\textwidth]{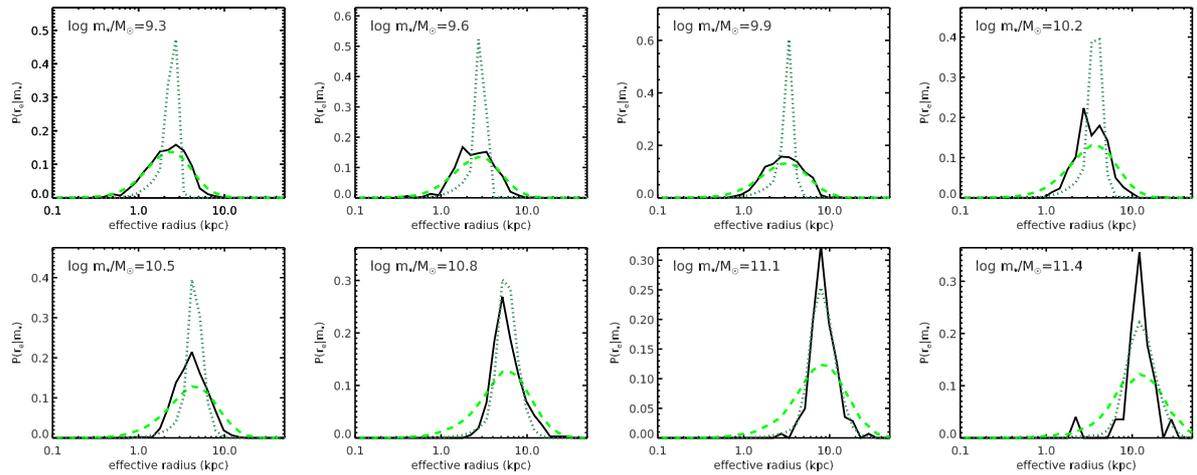}
\end{center}
\caption{Conditional probability distributions for effective radius in
  bins of stellar mass, at $z\sim 0.1$. Stellar mass bins increase
  from left to right and top to bottom, as indicated in the panel
  labels.  Black solid lines show the distribution of estimated 3D
  half-stellar mass radius ($r_{\rm 3D, *}$) from the GAMA
  observations.  Green dashed lines show distributions of SRHR$\lambda
  (\lambda R_h)$ from the SHAMs using a constant value of
  SRHR$\lambda=0.5$. Dark green dotted lines show distributions of
  SRHR $R_h$. This result places limits on the galaxy to galaxy
  dispersion in SRHR and SRHR$\lambda$.
\label{fig:sizedist_sham}} 
\end{figure*}

Fig.~\ref{fig:sizedist_sham} shows the conditional size distributions
$P(r_e|m_*)$ in stellar mass bins from GAMA (using the corrected 3D
stellar half mass radii), compared with the corresponding
distributions of $P(R_h|m_*)$ and $P(\lambda R_h|m_*)$ in stellar mass
bins from the SHAM. The distribution $P(\lambda R_h|m_*)$ in the SHAM
is very close to log-normal, as is well known to be the case for the
spin parameter in cosmological simulations
\citep[e.g.][]{bullock:2001b}.
In the lower stellar mass bins ($\log (m_*/M_\odot) \lesssim 10.8$,
the distribution of $P(R_h|m_*)$ is narrower than the observed
distribution $P(r_e|m_*)$, while the dispersion in the distribution
$P(\lambda R_h|m_*)$ matches the observed dispersion in $P(r_e|m_*)$
quite well. In the higher stellar mass bins, the dispersion in
$P(R_h|m_*)$ is already as large as the dispersion in $P(r_e|m_*)$,
while the dispersion in $P(\lambda R_h|m_*)$ is larger than the
observed dispersion. We discuss the interpretation of this result
further in \S\ref{sec:sizedisp}.
  
\subsection{Results from the CANDELS survey at $0.1<z<3.0$}

\begin{figure*} 
\begin{center}
\includegraphics[width=\textwidth]{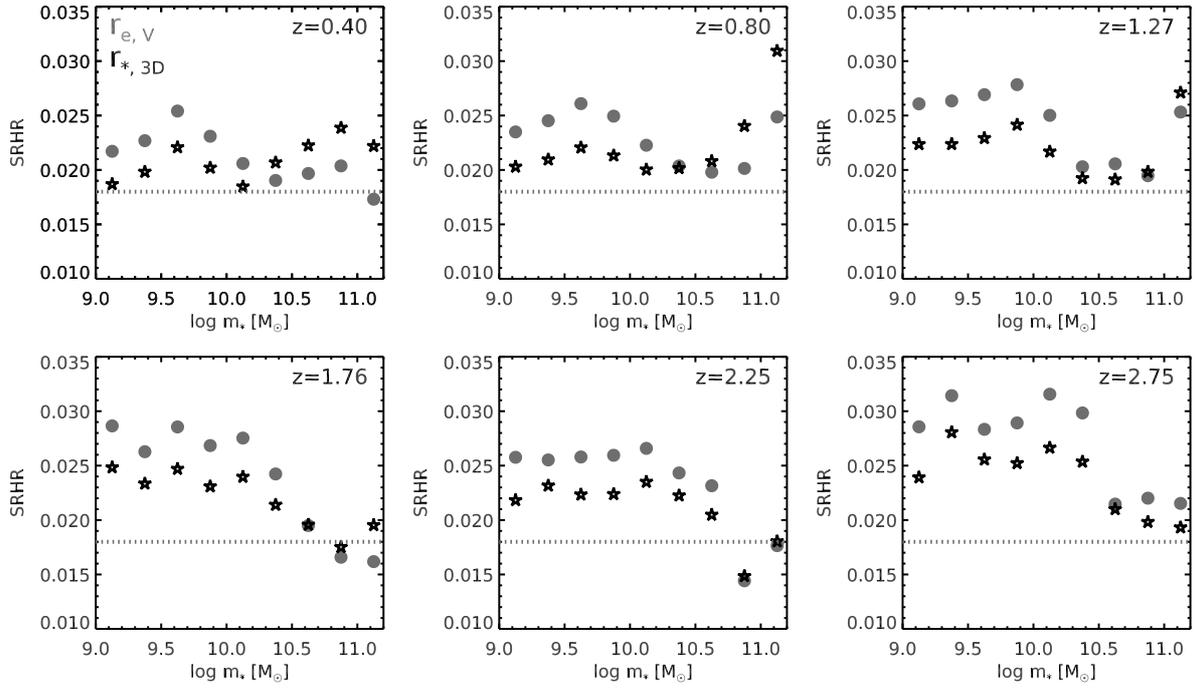}
\end{center}
\caption{The median observed radius in a stellar mass bin divided by
  the median value of halo radius $R_h$, from $z\sim 0.1$--3 (the
  values indicated in each panel are the volume midpoints of each
  bin).  Filled circles (gray) and stars (black) show results for the
  observed (projected) half-light radius $r_e$ and the 3D half-stellar
  mass radius ($r_{\rm 3D, *}$).  The dotted horizontal gray line
  shows the average $z=0.1$ value of SRHR from our analysis of the
  GAMA survey. SRHR has a stronger dependence on stellar mass in the
  higher redshift bins, and we see hints of a mild decrease of SRHR
  with cosmic time.
\label{fig:rgrh_ev}} 
\end{figure*}

We now investigate constraints on the dependence of SRHR and
SRHR$\lambda$ on stellar mass at different cosmic epochs. To do this,
we use the same SHAM approach, applied to mock CANDELS lightcones
extracted from the Bolshoi-Planck simulation.  We consider redshift
bins $0.1$--$0.5$, $0.5$--$1.0$, $1.0$--$1.5$, $1.5$--$2.0$,
$2.0$--$2.5$, and $2.5$--$3.0$. Fig.~\ref{fig:rgrh_ev} shows SRHR, the
ratio of median observed effective radius $r_e$ or de-projected half
stellar mass radius $r_{\rm *,3D}$ to the median value of $R_h$ in
stellar mass bins, for these six redshift bins. At $z=0.1$, we found
that SRHR is nearly constant across the full range of stellar masses
considered in our analysis. However, SRHR seems to gain a stronger
dependence on stellar mass as we move towards $z\sim3$, with more
massive galaxies having lower values.

\begin{figure*} 
\begin{center}
\includegraphics[width=\textwidth]{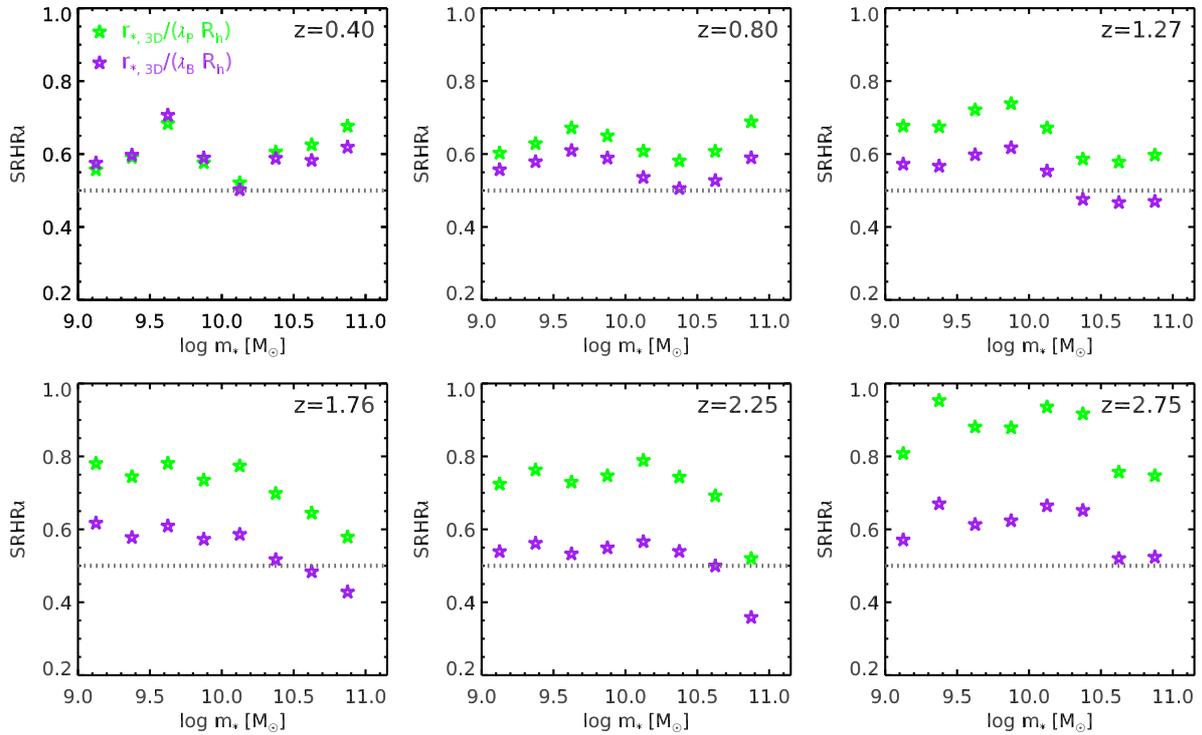}
\end{center}
\caption{The median observed radius in a stellar mass bin divided by
  the median value of halo radius $R_h$ times halo spin
  (SRHR$\lambda$), from $z\sim 0.1$--3 (the values indicated in each
  panel are the volume midpoints of each bin).  Here we use the 3D
  half-stellar mass radius ($r_{\rm 3D, *}$). Green symbols show the
  results using the Peebles spin definition, and purple show the
  results using the Bullock spin. The dotted horizontal gray line
  shows the average $z=0.1$ value of SRHR$\lambda$ from our analysis of the
  GAMA survey.
\label{fig:rgrhlambda_ev}} 
\end{figure*}

\begin{figure*} 
  \begin{center}
    \includegraphics[width=0.95\textwidth]{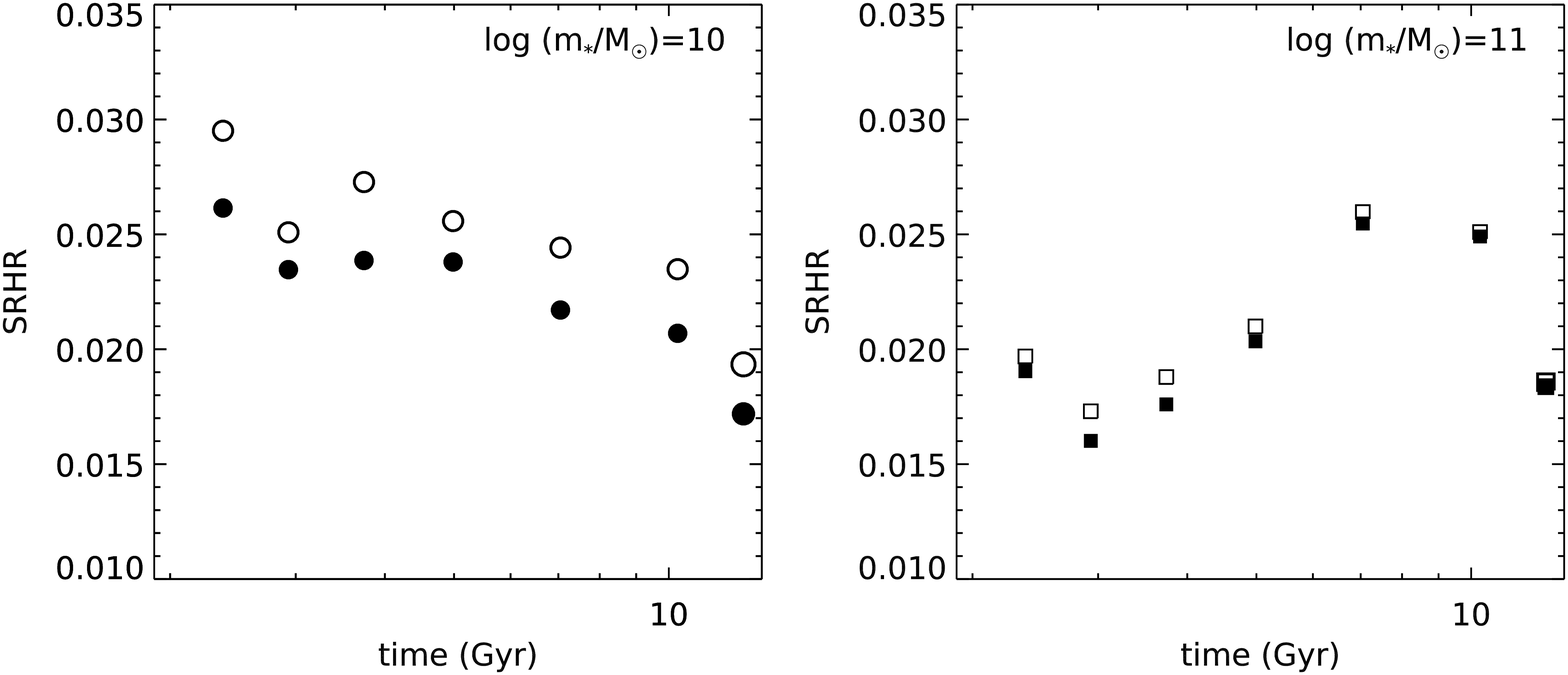}
    \includegraphics[width=0.95\textwidth]{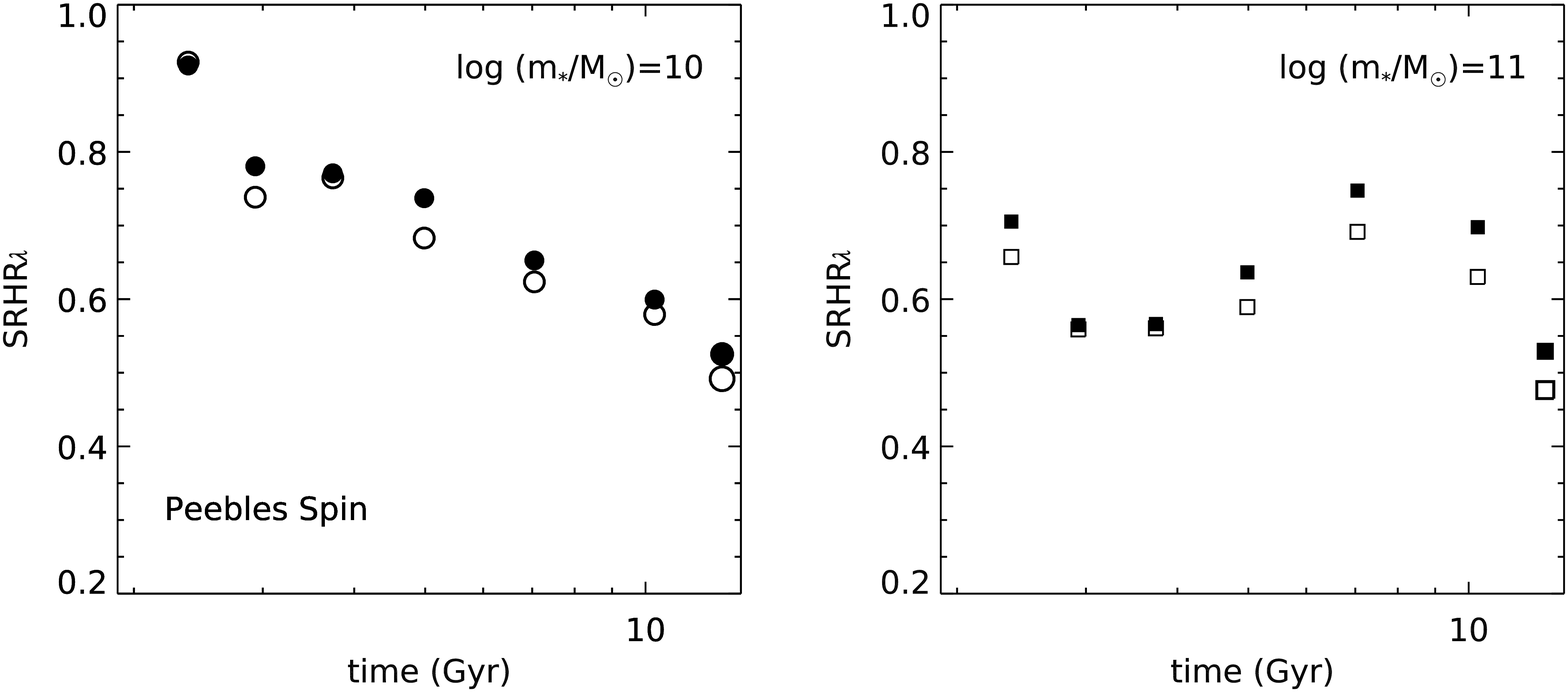}
    \includegraphics[width=0.95\textwidth]{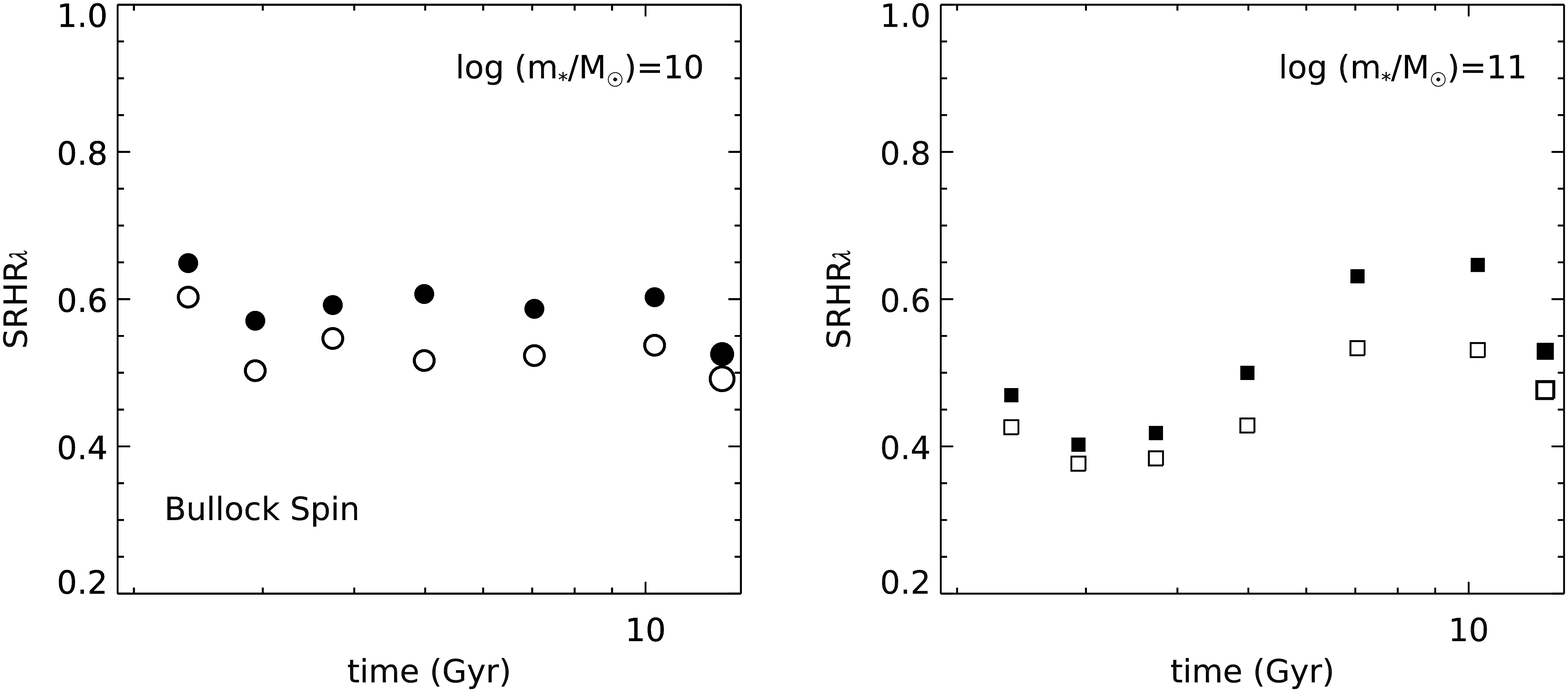}
  \end{center}
\caption{Time evolution of SRHR and SRHR$\lambda$, the ratio between
  median $r_{\rm *,3D}$ and $R_h$ or $\lambda R_h$, for two different
  stellar mass bins: $10^{9.75}\, \msun < m_* < 10^{10.25}\, \msun$
  (left; filled) and $10^{10.75}\, \msun < m_* < 10^{11.25}\, \msun$
  (right; filled). Top row: SRHR. Middle row: SRHR$\lambda$ using the
  Peebles spin definition; Bottom row: SRHR$\lambda$ using Bullock
  spin.  The ratio of the mean quantities is shown by the open symbols
  --- using means instead of medians results in slightly different
  numerical values, but does not change any of the trends.
\label{fig:frevtime}}
\end{figure*}

Fig.~\ref{fig:rgrhlambda_ev} shows SRHR$\lambda$, the ratio of median
de-projected stellar mass weighted radius $r_{\rm *,3D}$ to the median
value of $\lambda R_h$ in stellar mass bins, for six redshift bins as
before. We show results for both definitions of spin parameter
(Peebles and Bullock). We can see that any conclusions about the
evolution of SRHR$\lambda$ depend to a significant degree on which
spin definition is adopted, although trends with stellar mass are not
affected. This is because in the Bolshoi-Planck simulations, the
Peebles and Bullock spin parameters evolve differently with cosmic
time (as shown by \citet{rodriguez-puebla:2016} and in
Appendix~\ref{sec:appendix:halos} of this paper). We discuss the
possible reasons for the different evolution of the two definitions of
spin parameter in \S\ref{sec:spindef} and in the Appendix.

In Fig.~\ref{fig:frevtime}, we plot our results for SRHR and
SRHR$\lambda$ for two bins in stellar mass $\sim 10^{10} \msun$ and
$\sim 10^{11} \msun$ as a function of cosmic time since the Big Bang.
We again show our results for SRHR$\lambda$ for both definitions of
the spin parameter (Peebles and Bullock). For galaxies with $m_*
\lesssim 10^{10.5} \msun$, when using the Peebles spin, SRHR$\lambda$
declines by about a factor of 1.8 over the redshift range of our
study. Using the Bullock definition of spin, SRHR$\lambda$ in this
mass range is consistent with being constant in time.  For more massive
galaxies, SRHR$\lambda$ is nearly constant, or increases slightly,
over cosmic time within the CANDELS sample. The value of SRHR and
SRHR$\lambda$ derived from GAMA at $z\sim 0.1$ is about a factor of
1.4 lower than the CANDELS results from the lowest redshift bin. This
suggests that there may be a systematic offset between the size or
stellar mass estimates in GAMA and CANDELS for massive galaxies.

\begin{figure*} 
\begin{center}
\includegraphics[width=\textwidth]{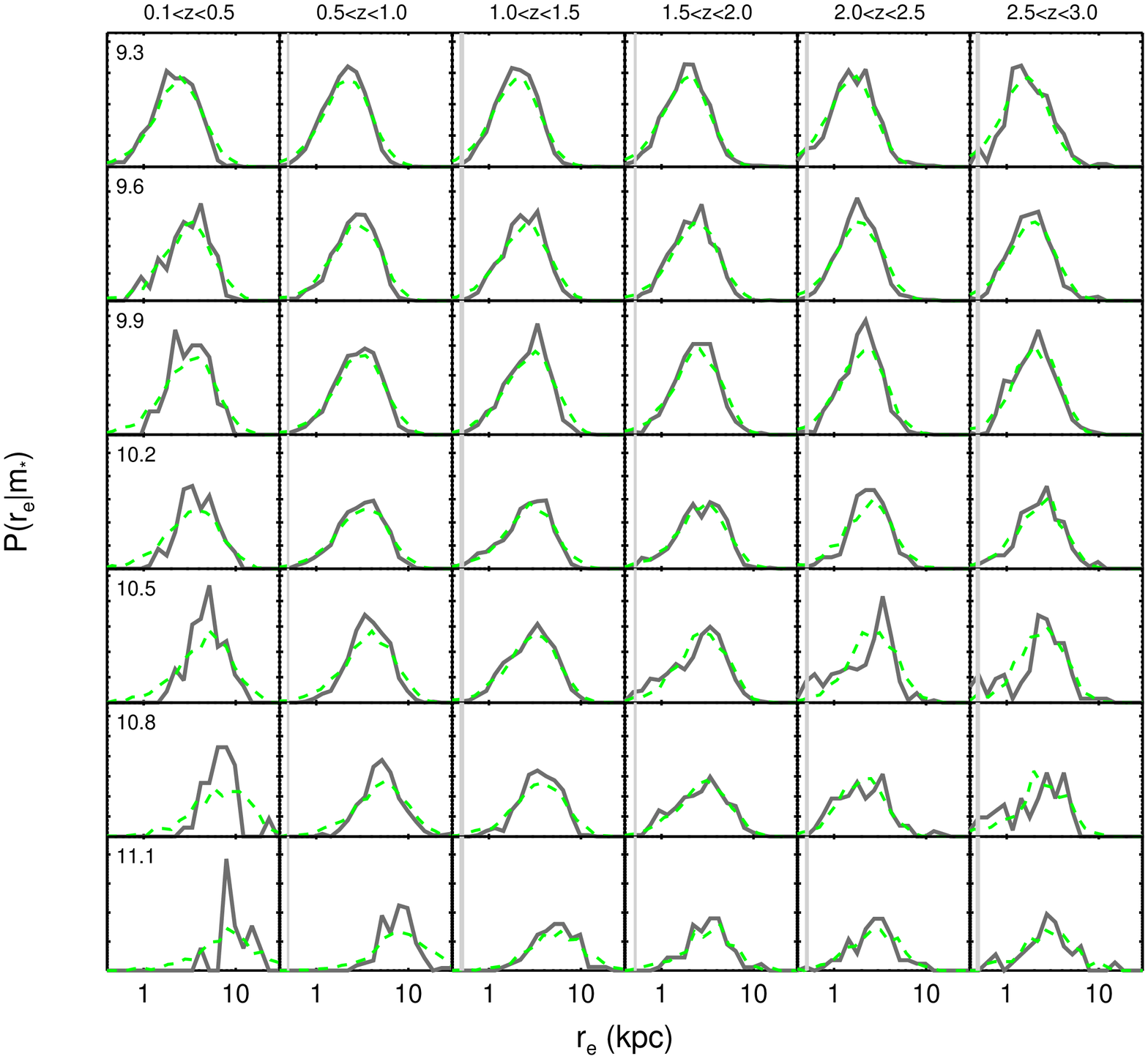}
\end{center}
\caption{The conditional probability distribution for effective radius
  $r_e$ in bins of stellar mass and redshift.  Gray solid lines show
  the distributions of the estimated 3D half-stellar mass radii
  ($r_{\rm *, 3D}$) from CANDELS. Vertical gray lines show the
  physical size corresponding to one F160W pixel in the drizzled image
  (0.06 arcsec), at the lower and upper limit of the redshift
  bin. Green dashed lines show distributions $P(\lambda R_h|m_*)$ from
  the SHAM (using the Peebles spin definition). The SHAM distributions
  have been shifted horizontally to match the medians of the observed
  distributions, to emphasize the comparison of the shapes of the
  distribution functions.
\label{fig:sizedistev}}
\end{figure*}

We now investigate the conditional distributions of galaxy radius in
stellar mass bins out to $z\sim 3$. Fig.~\ref{fig:sizedistev} shows
the conditional size distributions in bins of stellar mass and
redshift from CANDELS. Note that this diagram is similar to Fig. 10 of
vdW14, and the results appear similar, although here we show the
estimated deprojected stellar half-mass radii rather than the
projected half-light radii. We also show conditional distributions of
the quantity SRHR$\lambda \times (\lambda R_h)$ from the SHAM, where
we have used the redshift and stellar mass dependent values of
SRHR$\lambda$ derived above, and shown in
Fig.~\ref{fig:rgrhlambda_ev}, to shift the median of the SHAM
distribution to match that of the observed distributions.  In this way
we can compare the shape and width of the distributions in detail.  As
before, the SHAM distributions and their dispersions match the
observed conditional size distributions remarkably well. We show here
the results for the Peebles definition of the spin parameter, but the
distribution shapes are very similar for the Bullock definition. We
discuss the significance of these results in \S\ref{sec:sizedisp}.

\section{Discussion}
\label{sec:discussion}

In this section we discuss the main caveats and uncertainties in our
analysis, compare the ``forward'' and ``backwards'' modeling
approaches, compare our results and conclusions with those of previous
studies, and discuss possible physical interpretations of our
results. Some readers may wish to skip directly to
Section~\ref{sec:physics} for the discussion of the physical
interpretation of our results.

\subsection{Main Caveats and Uncertainties}

Our analysis makes use of, on the one hand, observational estimates of
galaxy stellar mass, redshift, and radial size (and, secondarily,
morphological type), and on the other, predictions of the mass,
radius, and spin parameters of dark matter halos from a cosmological
simulation.

\subsubsection{Halo properties and SMHM relation}

There are several important caveats to note regarding the halo
properties and SMHM relation. First, the halo masses, virial radii,
and spin parameters are taken from dissipationless N-body simulations,
which do not include the effect of baryons on halo properties. Studies
that do include baryons and the associated feedback effects have shown
that baryonic processes can modify the virial mass and spin parameter
of dark matter halos by up to 30\% \citep{munshi:2013,teklu:2015} and
the magnitude of these effects may depend on halo mass. Therefore the
\emph{actual} ratio of galaxy size to halo size and spin parameter may
differ from the values quoted here.

Second, specific properties of dark matter halos such as mass, radius,
and spin parameter depend on the definition used. See
Appendix~\ref{sec:appendix:halos} for a detailed description and
illustration of different halo mass, radius, and spin definitions.

How would our results change had we adopted a different halo
definition? The halo definition impacts several aspects of our
calculation. Recall that we have used the definition $M_{\rm vir,
  crit}$ as given in Section~\ref{sec:model}. Halos with a fixed value
of $M_{\rm 200,crit}$ are less abundant (have a lower volume density)
than halos with the same numerical value of $M_{\rm vir,
  crit}$. Similarly, halos with a fixed value of $M_{\rm vir, crit}$
are less abundant than halos with the same numerical value of $M_{\rm
  200,b}$. This means that galaxies with a given stellar mass (and
observed number density) will be assigned larger and larger halo
masses depending on the halo definition used, from $M_{\rm 200,crit}
\rightarrow M_{\rm vir, crit} \rightarrow M_{\rm 200,b} \rightarrow
M_{\rm vir, b}$. Moreover, the virial radius for a given halo mass
increases as we go from $M_{\rm 200,crit} \rightarrow M_{\rm vir,
  crit} \rightarrow M_{\rm 200,b} \rightarrow M_{\rm vir, b}$. Since
$r_e$ for a given $m_*$ is fixed by the observed relation, all of this
implies that $r_e/R_{h}$ would be largest for the $M_{\rm 200,crit}$
definition and smallest for the $M_{\rm vir, b}$ definition. Our
favored definition is in the middle. Furthermore, we expect $\lambda$
to increase slightly as we go from $M_{\rm 200,crit} \rightarrow
M_{\rm vir, crit} \rightarrow M_{\rm 200,b} \rightarrow M_{\rm vir,
  b}$. This means the difference in $r_e/(\lambda R_h)$ will be even a
bit larger from one halo definition to another. To accurately fully
estimate the effects of changing the halo definition, we would need to
redo the abundance matching and remeasure $\lambda$ consistently for
each definition, which is beyond the scope of this paper. However, a
crucial point is that we have been very careful to use a consistent
halo mass definition in \emph{all} aspects of our study.

The choice of halo definition is in some sense arbitrary. Yet, one can
ask which definition is the most physically relevant for tracking
quantities that are relevant to galaxy formation, such as the
accretion rate of gas into the halo. Some recent works that examined
structure formation in dark-matter only simulations have pointed out
that defining the halo relative to an evolving background density
leads to apparent growth of the halo mass even as the physical density
profile of the interior of the halo remains unchanged, an effect that
has been termed ``pseudo-evolution''
\citep{busha:2005,diemer:2013}. This suggests that this mass growth
should not be associated with physical accretion of matter into the
halo. However, some more recent studies that have examined simulations
including baryonic physics find that the accretion rate of \emph{gas}
into the central part of halos (onto forming galaxies) tracks the
growth of the virial radius quite well
\citep{dekel:2013,wetzel:2015}. This implies that while
pseudo-evolution is a relevant concept for dark matter, not so for
baryons, which can shock and cool. The work of both \citet{dekel:2013}
and \citet{SHARC:2016} support the physical relevance of the halo mass
definition adopted here, and it is quite similar to the 200 times
background definition that was found to trace gas accretion by
\citet{wetzel:2015}.

Another important caveat is the adopted SMHM relation, which plays a
critical role in our analysis. As already noted, the abundance of dark
matter halos as a function of their mass depends on the halo mass
definition, but it also depends on the method used to identify halos
and sub-halos in the N-body simulation. The cumulative halo mass
function at $z=0$ differed by $\pm 10$\% across the 16 halo finders
tested in \citet{Knebe:2011}. However, much larger differences between
different halo finders can arise at high redshift
\citep{Klypin:2011}. Phase-space based methods such as the ROCKSTAR
halo finder used here tend to be the most robust
\citep{Knebe:2011}. In deriving the SMHM relation, there are also
subtleties regarding how sub-halos are treated: whether to use their
properties at infall or at the time they are identified (these can
differ substantially due to tidal stripping), and whether
sub-halos/satellites obey the same SMHM relation as central
galaxies. As noted above, sub-halos/satellites should be sufficiently
sub-dominant in our sample that these details will not have a large
impact on our results.

The main reason for systematic differences between SMHM relations
quoted in the literature is in fact the lack of convergence between
different observational determinations of the stellar mass
function. This is most acute for very massive nearby galaxies
\citep{kravtsov:2014,bernardi:2013} and also at high redshift -- even
at $z\sim 1$ there is a lack of convergence regarding the low-mass
slope of the stellar mass function \citep[see
  e.g.][]{moster:2010}. The SMHM relation used in this work is based
on a very recent and complete compilation of stellar mass functions,
and adopts the same cosmological parameters as in our work. We have
confirmed that when we apply the RP17 SMHM relation with our adopted
scatter to our halo catalogs, we reproduce the GAMA stellar mass
function at $z=0.1$ and the CANDELS stellar mass functions at $z\sim
0.1$--3. As we showed in Fig.~\ref{fig:smhm}, several recent
determinations of the SMHM relation are in good agreement over the
mass range relevant to our study ($10^{9} \lesssim m_* \lesssim
10^{11} \msun$). We further note that the differences between the RP17
and B17 $z=0.1$ SMHM relation seen in Fig.~\ref{fig:smhm} do not
significantly affect our results, because we focus on galaxies less
massive than a few $10^{11} \msun$, where the differences are
small. However, adopting a larger scatter in the SMHM relation leads
to larger values of SRHR at high stellar masses ($m_* \gtrsim
10^{10.5}$), as more galaxies hosted by lower mass halos are scattered
into these bins.

\subsubsection{Definition of halo spin parameter}
\label{sec:spindef}

We have seen that our results for the evolution of SRHR$\lambda$ are
quite different for two commonly used definitions of the spin
parameter $\lambda$.  As shown by \citet{rodriguez-puebla:2016}, at
$z=0$ the distributions of $\lambda_B$ and $\lambda_P$ peak at nearly
the same value, but $\lambda_B$ has a more pronounced tail to larger
values (see Fig.~21 of \citet{rodriguez-puebla:2016}).  However,
$\lambda_P$ increases from $z\sim 3$--0, while $\lambda_B$ decreases
slightly over this interval. Very similar results are shown in a
recent analysis of the Illustris simulations by
\citet{zjupa:2017}. This explains why we found milder evolution in
SRHR$\lambda$ when using the Bullock definition $\lambda_B$.

We discuss possible reasons for the different behavior of $\lambda_P$
and $\lambda_B$ in Appendix~\ref{sec:appendix:halos}. We conclude that
this is likely due to a combination of changing halo density profiles
(concentration), deviation of halos from perfect spheres, and/or
changes in halo kinematics (deviation from circular orbits). This
brings up further concerns regarding the basis of simple analytic
models of disk formation, which assume that all halo particles are on
circular orbits.

Which definition of halo spin is more physically relevant to the
question at hand, namely galaxy sizes? We feel that this is not
currently clear. In some sense, the Peebles definition appears to
capture some real and potentially relevant evolution in halo structure
and kinematics. Moreover, it is the Peebles definition of $\lambda$
that properly comes in to somewhat more sophisticated analytic models
of disk formation \citep[e.g.][]{mo:1998,somerville:2008a},
i.e. Eqn.~\ref{eqn:mmw} below.  \citet{zjupa:2017} find that the
Peebles definition is more robust than the Bullock definition for
halos defined by the friends-of-friends (FOF) method. The relevance of
either quantity to observed galaxy sizes should be explored further
using detailed numerical simulations of galaxy formation, but the
potentially significant differences between these two definitions
should be kept in mind.

\subsubsection{Observational measurements}

Analogous to the problem of defining a halo, there is no unique way to
define the total amount of light within a galaxy, as galaxies do not
have sharp edges. This necessarily leads to an ambiguity in how the
half-light radius is defined, as it is defined relative to the total
amount of light. Commonly used metrics include isophotal magnitudes
(and sizes), Petrosian or Kron magnitudes and sizes, model magnitudes
and sizes, and the curve-of-growth method
\citep{bernardi:2014,curtis-lake:2016}. Here we have used model sizes,
where the model is a single component S{\'e}rsic profile. Some
galaxies are not well-fit by a single component S{\'e}rsic profile,
and one might expect our method to do poorly in these cases. In the
local universe, the largest discrepancy in total luminosity, stellar
mass, and size is for very massive giant elliptical galaxies
\citep{bernardi:2013,bernardi:2014}. Our CANDELS sample is dominated
by lower mass galaxies, so that part of our analysis should not be
greatly affected by these objects.  Model fitting based sizes can also
be sensitive to the local background used in the fitting, and to the
seeing or point spread function (PSF) of the image. We adopted the
GAMA sample for our study because the methods used to estimate stellar
masses and sizes were as similar to those used for CANDELS as any
low-redshift sample of which we are aware. In both GAMA and CANDELS,
sizes are estimated using the same code (GALFIT) and single component
S{\'e}rsic fitting.

Another important note is that some studies
\citep[e.g.][]{shen:2003,shibuya:2015} have used \emph{circularized}
radii ($r_{\rm e, circ} \equiv q^{1/2}\, r_{\rm e, major} $ where $q$
is the projected axis ratio), rather than semi-major axis
radii. Because galaxy axial ratios can depend on stellar mass and
redshift, this could lead to different conclusions.

Further uncertainties come from the conversion from observed,
projected (2D) radii to physical 3D radii, which depends on the shape
of the galaxy (flat versus spheroidal). Again, this is probably
correlated with stellar mass and may vary with cosmic time. We have
attempted to make a crude correction for these dependencies but this
should be improved. In a similar vein, we used the empirical
corrections of vdW14 to correct from observed-frame H$_{160}$ size to
rest-frame 5000\AA\ size, and then further attempted to convert from
observed rest-frame 5000\AA\ half-light radius to stellar half-mass
radius. The mass, redshift, and type dependences of these corrections
also remain uncertain and poorly constrained. It should be possible to
better account for this in the future by doing pixel-by-pixel SED
fitting to measure stellar mass profiles \citep{wuyts:2012}.

\subsection{Beware Backwards Modeling}
\label{sec:bewareback}
\begin{figure*} 
\begin{center}
\includegraphics[width=\textwidth]{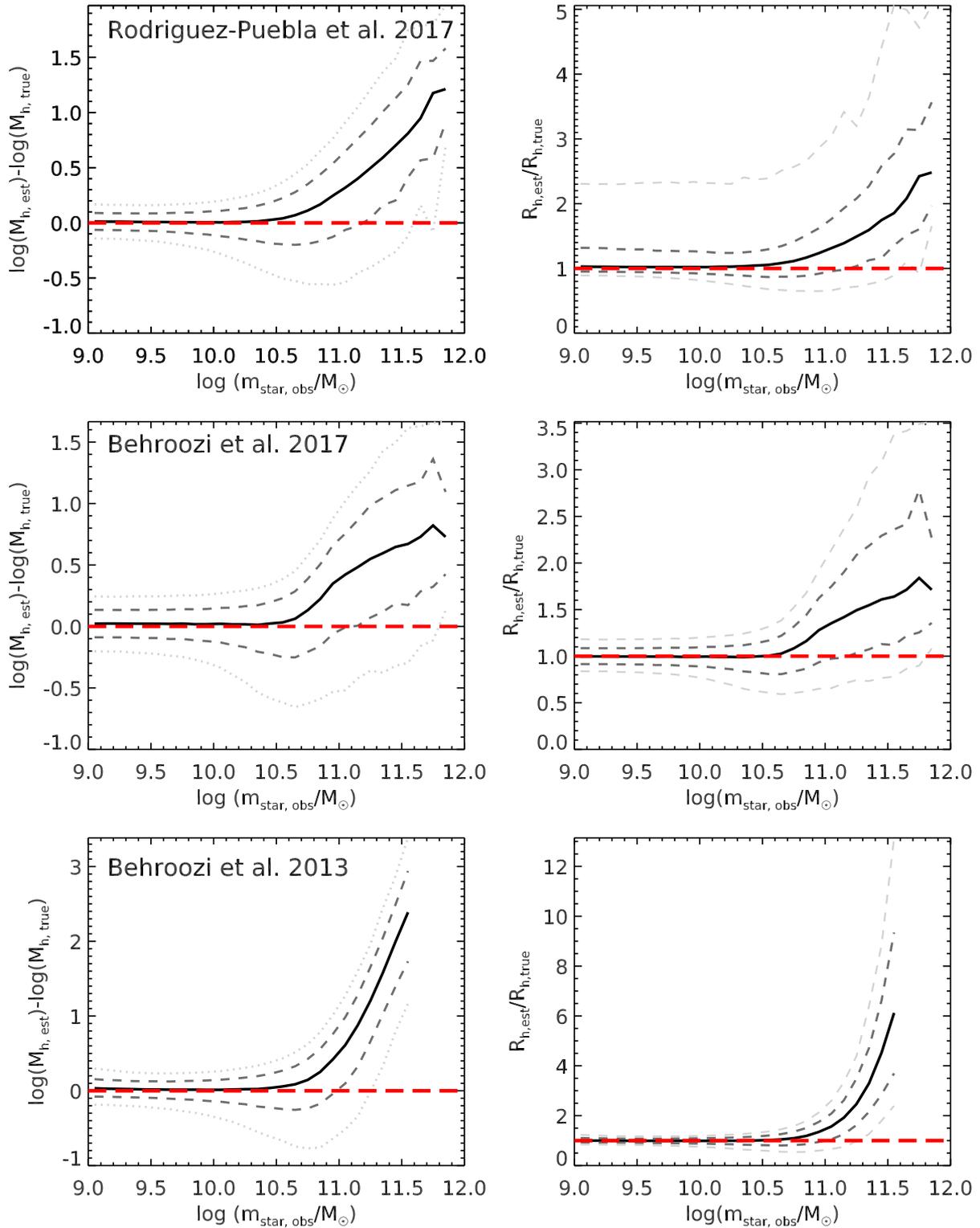}
\end{center}
\caption{A test of the accuracy of recovering halo mass and radius
  estimates from backwards modeling, based on applying this method to
  a mock catalog in which the true halo masses and radii are
  known. Different panels show the results from mock catalogs created
  with different SHAM models, as indicated on the panels. The left
  column shows the difference between the log of the halo mass
  estimated by backwards modeling and the log of the true halo mass,
  and the right panels show the ratio of estimated to true halo virial
  radius. Red lines indicate equality, solid black lines show the
  medians, medium gray dashed lines show the 16th and 84th
  percentiles, and light gray dashed lines show the 2nd and 98th
  percentiles.  The median recovered halo mass and radius is fairly
  accurate below the ``turnover'' in the SMHM relation slope, but
  above this critical mass, the errors can become very large. The
  error in recovering the halo mass depends on the slope of the SMHM
  relation and the scatter in the SMHM relations due to intrinsic
  dispersion and stellar mass errors.
\label{fig:backtest}} 
\end{figure*}

In the approach used here, we start from an ensemble of dark matter
halos and sub-halos from theoretical cosmological simulations, and
apply empirical relations to map halo mass to stellar mass. We refer
to this approach as ``forward modeling''. An alternative approach,
sometimes used in the literature, is what we refer to as ``backwards
modeling''. In backwards modeling, halo masses and radii are derived
for an observational sample based on a stellar mass estimate. This is
often done by inverting a SMHM relation $\langle m_*(M_h)\rangle$
derived from abundance matching. However, this practice can be quite
dangerous in the presence of scatter in the underlying SMHM relation,
as we now show. From Fig.~\ref{fig:smhm}, we can see that above a
characteristic value of $M_h$, the slope of the SMHM relation becomes
quite shallow. As a result, a positive deviation in stellar mass
$\Delta m_*$ leads to a larger deviation in the derived halo mass than
a corresponding negative $\Delta m_*$, leading to a systematic
overestimate in halo mass and radius. Moreover, due to Eddington bias,
as stellar mass increases above the ``knee'' in the stellar mass
function, an increasing fraction of galaxies with estimated stellar
masses in a given stellar mass bin are likely to have been scattered
there due to stellar mass errors.

Fig.~\ref{fig:backtest} shows a test based on applying backwards
modeling (inversion of a SMHM relation) to a mock catalog in which the
true halo properties are known. We create a mock catalog of stellar
masses based on the Bolshoi-Planck simulation, by applying an assumed
SMHM relation with a log-normal scatter in stellar mass at fixed halo
mass, as described in Section~\ref{sec:model}. Our mock catalog
reproduces the observed stellar mass function at $z=0.1$. Below the
mass scale where the SMHM becomes shallower, the median recovered halo
properties are nearly unbiased. However, above this mass scale ($m_*
\simeq 10^{10.5} \msun$), the median recovered halo mass can be
overestimated by as much as two orders of magnitude, and the estimated
median halo radius can be overestimated by up to a factor of six. We
show this test at $z\sim 0.1$ as an illustration, but in detail the
errors in recovered parameters will depend on the stellar mass errors
and the slope of the SMHM relation.

\subsection{Comparison with Previous Work}

\begin{figure*} 
\begin{center}
\includegraphics[width=\textwidth]{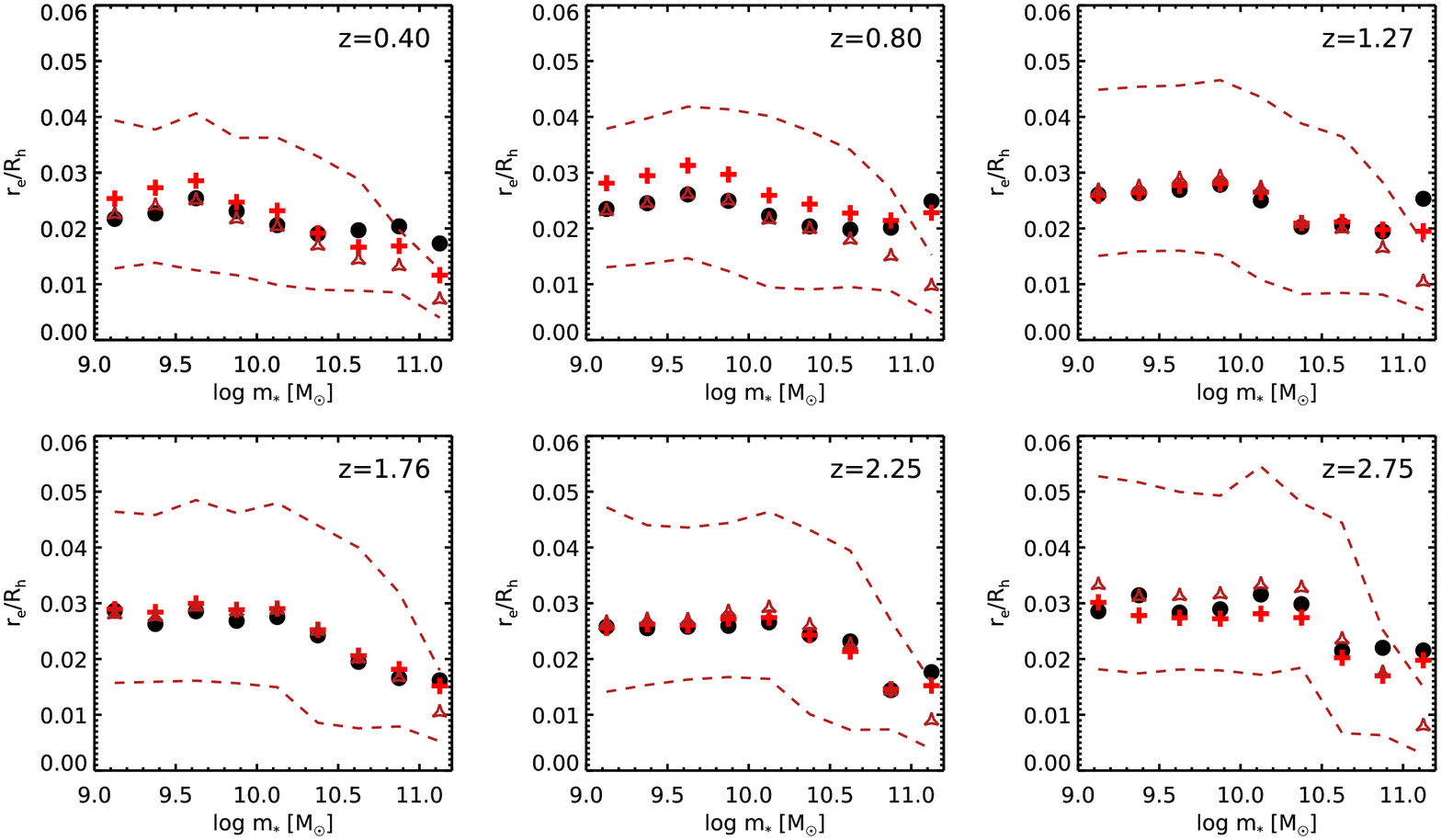}
\end{center}
\caption{Ratio of (observed, projected) galaxy effective (half-light)
  radius to halo radius as a function of stellar mass and
  redshift. Filled circles show the analysis presented in this
  work. Red crosses show the median values of $r_e/R_h$ from the
  analysis of H17, using their fiducial halo mass definition and SMHM
  relation (which are different from ours). Dark red triangles show
  results from the approach of H17 using our halo definition and the
  \protect\citet{behroozi:2013} SMHM relation. The dashed lines show
  the 16 and 84th percentiles from H17, which estimates $R_h$ for
  individual galaxies using the ``backwards modeling'' approach. Our
  results are in excellent agreement, except at the highest stellar
  masses, where the differences are likely due to the forward
  vs. backwards modeling approach (see text).
\label{fig:fr_ev_kuang}} 
\end{figure*}

Our $z\sim 0.1$ analysis of GAMA yields results that are very similar
to those of the analysis of K13, which was based on a more
heterogeneous low redshift sample that spanned a larger range in stellar
mass. Interestingly, in spite of the fact that we used different
observational samples and different halo mass definitions, our
quantitative conclusions for the low-redshift part of the analysis are
very consistent with those of K13: galaxy effective radius is
linearly proportional to halo radius with a proportionality factor of
$\sim 0.018$ (K13 finds 0.015). However, our physical interpretation
of the results is quite different from that of K13, as we discuss
further below.

Another recent study that has examined the relationship between galaxy
size and halo size using abundance matching is that presented in two
papers, \citet{shibuya:2015} and \citet{kawamata:2015}. They also
analyze the CANDELS+3D-HST sample as well as an additional sample of
Lyman-break galaxies. They perform their own GALFIT fitting procedure
to measure the sizes of the CANDELS+3D-HST sample as well as the LBG
sample. They find good statistical agreement between their measured
sizes and those of vdW14 for the CANDELS+3D-HST
sample.
They then estimate the dark matter halo radius for each galaxy based
on its stellar mass, using the abundance matching relation of
\citet{behroozi:2013}, and use this to estimate $r_e/R_h$. They find
values of $r_e/R_h = 0.01-0.035$, with ``no strong evolution'' in
$r_e/R_h$ from $z\sim 0$--8. This is broadly consistent with our
results. However, there are several differences between their analysis
and ours, which make our results difficult to compare in detail.  For
their main analysis (for which they compute $r_e/R_h$), galaxies are
selected in bins of observed UV luminosity, rather than stellar mass,
and sizes are k-corrected to the UV rather than the rest-frame
optical. They use a different halo mass definition than we do, and
indeed than \citet{behroozi:2013}. They use circularized radii, which
as we have noted may have a redshift-dependent relationship with
semi-major axis radii. If one looks closely at their Fig.~16,
focussing on the $z\lesssim 3$ redshift range of our study, there is a
difference of almost a factor of two between different bins in UV
luminosity at fixed redshift. Assuming that UV luminosity roughly
traces SFR, it is well-known that there is a declining relation
between stellar mass and SFR with decreasing redshift \citep[e.g.][and
  references therein]{speagle:2014}. One might expect, then, that
selecting galaxies at a fixed star formation rate would select lower
mass galaxies at high redshift. Furthermore, even at fixed stellar
mass there is a correlation between size and SFR, such that galaxies
with below-average SFR for their epoch have smaller sizes
\citep{wuyts:2011,brennan:2017}.
Finally, as noted by \citet{behroozi:2013}, the inverse of the fitting
formula for the average stellar mass at a given halo mass is not
equivalent to the average halo mass at a given stellar mass, because
of scatter in the stellar-mass-halo-mass
relation. \citet{shibuya:2015} ``backward'' model (go from stellar
mass to halo mass) while we ``forward'' model (go from halo mass to
stellar mass).

The recent study of H17 is easier to compare with our results, as they
use the same CANDELS catalogs and size measurements used in our
study. H17 perform a slightly different sample selection from the
parent CANDELS catalogs than we do. While we apply a uniform magnitude
limit that is appropriate for the CANDELS wide depth (H$_{160}<24.5$),
H17 apply a fainter magnitude cut in the CANDELS deep and Hubble
Ultra-deep field (HUDF) regions. H17 demonstrate the important result
that the size distributions for objects in the magnitude range $23.5 <
H_{160} < 24.5$ in the wide region and HUDF are consistent, confirming
that low surface brightness objects or wings are not biasing the size
distributions significantly at these magnitudes. As we show in
Fig.~\ref{fig:sizemass_candels}, the size mass relations that we
derive are nearly identical to those obtained from the sample of H17.

In Fig.~\ref{fig:fr_ev_kuang}, we show a comparison between our
derived values of $r_e/R_h$, where $r_e$ is the observed (projected)
rest-frame 5000\AA\ half light radius and $R_h$ is the halo radius,
and the median values of $r_e/R_h$ from the analysis of H17. Overall,
the results are in excellent agreement, particularly when they repeat
their analysis using the same halo mass definition, and a similar SMHM
relation, as those adopted in our study. We see hints of a larger drop
in $r_e/R_h$ at the highest stellar masses, which may be because the
``backward modeling'' approach adopted by H17 can tend to overestimate
halo mass and radius in the presence of scatter in the SMHM relation
(see Section~\ref{sec:bewareback}). 

H17 show the $R_h$-$r_e$ relation separately for galaxies with the
lowest and highest values of S{\'e}rsic index and of specific star
formation rate (sSFR). We are unable to do this in our forward
modeling approach. H17 find that the lowest S{\'e}rsic (disky)
galaxies have larger values of $r_e/R_h$ than the highest S{\'e}rsic
galaxies. A similar result holds for the highest and lowest sSFR
galaxies (the highest sSFR galaxies have larger $r_e/R_h$). This is
consistent with our finding that SRHR is smaller for higher stellar
mass bins, which also tend to have larger fractions of
high-S{\'e}rsic, low-sSFR galaxies. However, we note that H17 have not
attempted to perform any correction for the different conversion
between projected and 3D radius for flat and round galaxies.

\subsection{Physical Interpretation}
\label{sec:physics}
\subsubsection{Theoretical expectations for disk sizes}

What do our results tell us about the physics that shapes galaxy
sizes? We first compare our results with the predictions of the
simplest model for disk formation, Eqn~\ref{eqn:rd_iso}. We can
re-write this as:
\begin{equation}
\frac{r_e}{\lambda R_h} = \frac{1.678}{\sqrt{2}}\, f_j
\end{equation}
where $f_j$ is the ratio of the specific angular momentum of the disk
to that of the halo. If this na\"{i}ve model were correct, then if the
specific angular momentum of the stellar disk is the same as that of
the halo ($f_j=1$), we would have $r_e/(\lambda R_h)= 1.18$. 

Several refinements to this simplest model have been presented in the
literature. First, dark matter halos that form in dissipationless
N-body simulations in the \LCDM\ paradigm do not have singular
isothermal density profiles, but are better characterized by the
Navarro-Frenk-White \citep[NFW;][]{navarro:1997} functional
form. Second, in the absence of non-gravitational energy injection,
self-gravity from the baryons that collect in the center of the dark
matter halo following cooling and dissipation should lead to
contraction, leading to disks that are smaller than the na\"{i}ve
model would predict. The degree of contraction can be estimated using
the ``adiabatic invariant'' approximation
\citep[e.g.][]{blumenthal:1986,flores:1993,mo:1998}. In this
formalism, the contraction factor depends on the halo concentration,
the disk mass fraction, and the halo spin parameter, where more
concentrated halos, heavier disks, and lower spin parameters lead to
more contraction \citep[see][]{dutton:2007,somerville:2008a}.

In these slightly more sophisticated models, we now obtain
\citep{mo:1998,somerville:2008a}:
\begin{equation}
\frac{r_e}{\lambda R_h} = \frac{1.678}{\sqrt{2}} f_j f_c^{-1/2} f_R (\lambda,c,f_d)
\label{eqn:mmw}
\end{equation}
where $c$ is the NFW concentration parameter and $f_d \equiv m_{\rm
  disk}/M_{h}$ is the baryonic mass of the disk in units of the total
halo mass. The functions $f_c$ and $f_R (\lambda,c,f_d)$ account for
the NFW profile and the adiabatic contraction. As shown in
\citet{somerville:2008a}, typical values for $f_c^{-1/2} f_R$ range
from 0.4 to unity, and likely have an effective dependence on redshift
through the evolving halo mass vs. halo concentration relationship
(see the extensive discussion in \citet{somerville:2008a}). Thus, in a
pure adiabatic contraction picture, ignoring the presence of gas, we
would have to conclude that $f_j$ must be unity or greater than unity.

However, the quantity $f_d$ that enters above is the total
\emph{baryonic} mass of the disk. In low mass and high redshift
galaxies, cold gas in the interstellar medium can comprise comparable
or even possibly greater amounts of mass than stars. Furthermore, the
size predicted by this equation is the size of the \emph{baryonic}
disk (stars plus cold gas). It is well known that atomic gas is much
more extended than the stellar disks in nearby galaxies
\citep{bigiel:2012}.  However, it is unknown how the stellar half-mass
radius tracks the total baryonic effective radius as a function of
mass and redshift. \citet{berry:2014} presented arguments based on
modeling of Damped Lyman-$\alpha$ systems that this ratio might have
to evolve with redshift. Due to these considerations and other
complications, we do not attempt to draw any strong conclusions about
$f_j$ from this work.

K13 points out that the normalization of the $r_e$ vs. $R_{200}$
relation implied by his analysis is about a factor of two lower than
that predicted by the simple disk formation model with adiabatic
contraction (Eqn.~\ref{eqn:mmw}). He speculates that this could be
because the galaxy size reflects the size of the halo when the disk
formed, rather than at the present day. He further speculates that
most of the apparent growth in halo mass and size since $z\sim 2$ is
due to ``pseudoevolution''. This would imply that galaxy growth does
not track the halo growth from $z\sim 2$--0, so the galaxy size should
be proportional to the halo's size at $z\sim 2$. We find that a more
detailed implementation of the standard adiabatic contraction model
within a full semi-analytic merger tree model (including the effects
of gas, disk instabilities, and mergers; Somerville et al. in prep)
produces disks that are about 50\% too large at a given mass at $z\sim
0$, compared with observations, but are in good agreement with the
size-mass relation from CANDELS at $0.4 \lesssim z \lesssim 3$
\citep[see also][]{brennan:2017}. However, we do not support
``pseudoevolution'' as a complete explanation for two reasons. First,
the concept of pseudoevolution does not appear to apply to gas within
forming halos, as discussed above \citep[see also arguments presented
  in][]{SHARC:2016}. Second, the ``stagnation'' of disks since $z\sim
2$ does not appear to be consistent with the star formation histories
of galaxies derived from multi-epoch abundance matching.

\begin{figure} 
\begin{center}
\includegraphics[width=0.5\textwidth]{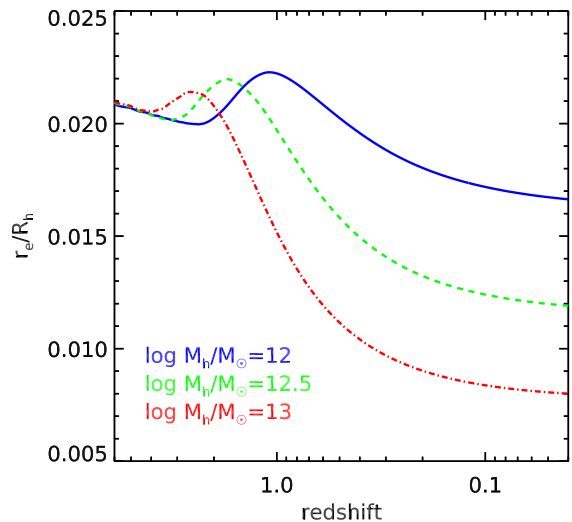}
\end{center}
\caption{Examining the effect of star formation history on galaxy
  size. Using halo-mass dependent star formation histories derived
  from abundance matching, we assume that new stars were formed in an
  exponential disk with a half-mass radius of $0.036\, R_h(z)$, with
  the same center and orientation as previous generations.  At each
  redshift, we stack all stars formed (accounting for appropriate
  stellar mass loss) and compute the ratio of the stellar half-mass
  radius to $R_h$.  For today's high-mass galaxies ($10^{13} M_\odot$
  halos), there is quite a lot of evolution in this ratio, because
  most of the stars were formed early on when the halo was much
  smaller, and there is little late star formation.  For lower-mass
  galaxies ($10^{12} M_\odot$ halos), the effect is smaller, about
  30\%.  For even lower masses, we expect the effect to be even
  smaller.
\label{fig:stellar_size_ratio}} 
\end{figure}

We illustrate this in Fig.~\ref{fig:stellar_size_ratio}. Here we use
the halo-mass dependent star formation histories derived from
abundance matching as described in \citet{behroozi:2013}. We assume
that stars were formed in an exponential disk with half-mass radius
$\langle \lambda \rangle R_h(z)$, where $\langle \lambda \rangle =
0.036$ is the average value of the spin parameter in Bolshoi-Planck,
and $R_h(z)$ is the halo virial radius at the redshift at which a
parcel of stars is formed. Fig.~\ref{fig:stellar_size_ratio} shows
that galaxies in massive halos $M_h \gtrsim 10^{12.5} M_\odot$ might
have sizes that more closely reflect the halo size in the past,
because star formation in these halos was quenched at some earlier
time. However, galaxies in halos the mass of our Milky Way or smaller
$M_h \lesssim 10^{12} M_\odot$ have had considerable ongoing star
formation, and therefore this ``formation time'' weighted size does
not change much. The evolution for even lower mass halos would be even
smaller.

Perhaps the only way to reconcile the idea that disk sizes reflect the
halo size at some earlier epoch with the results presented above would
be if the \emph{gas} stopped falling in to the disk at some point, and
star formation continued as that disk gas reservoir was converted into
stars. This probably happens to some extent. However, both numerical
hydrodynamic simulations
\citep{faucher-giguere:2011,angles-alcazar:2016} and observations of
galaxy gas content and consumption times at $z\sim 1$--2
\citep{Saintonge:2013,Genzel:2015} are inconsistent with gas accretion
ceasing completely at $z\sim 1$--2 in disk galaxies in Milky Way or
smaller sized halos.

\citet{desmond:2015} investigated a model based on abundance
  matching and an angular momentum partition model similar to
  Eqn.~\ref{eqn:mmw}, but allowing for expansion as well as
  contraction due to baryonic processes. They found that they were
  able to reproduce the normalization and slope of the $z\sim0$
  mass-size relation with a value of $f_j$ (in our notation) of 0.74
  to 0.87, depending on the halo property used in the abundance
  matching.

\subsubsection{Theoretical expectations for spheroid sizes}
One of the surprising results of this work (already emphasized by K13)
is that the linear proportionality $r_e \propto \lambda R_h$ seems to
work just as well for spheroid dominated galaxies at $z\sim 0.1$ as it
does for disks --- and in the nearby universe, the value of SRHR is
almost the same for stellar mass bins that are mostly comprised of
spheroid-dominated galaxies and those that are mostly comprised of
disk-dominated galaxies. However, a new result shown in this work is
that there are hints that SRHR has a stronger dependence on stellar
mass at high redshift, such that SRHR is smaller for higher mass
galaxies at high redshift. Why should SRHR be lower for high mass
galaxies at high redshift, but then converge to the same value as for
lower mass galaxies by $z\sim 0.4$?

This may be consistent with the picture in which massive galaxies at
high redshift experience significant loss of angular momentum and
compaction due to dissipational processes such as mergers and violent
disk instabilities
\citep{porter:2014,Dekel_Burkert:2014,Zolotov:2015}.  Some of the gas
that has been stripped of angular momentum is able to accrete onto a
central supermassive black hole, which subsequently drives gas out of
the galaxy with powerful winds and stops further cooling. The remnants
are then ``puffed up'' by dry (gas poor), mostly minor mergers
\citep{naab:2009,shankar:2010,shankar:2013,Hilz:2013,porter:2014,Porter2014b}.
As the galaxy then acquires angular momentum from the orbits of the
merged satellites, it is perhaps not so surprising after all that the
angular momentum of the satellite population traces that of the host
dark matter halo. We work these ideas out in more detail, and explore
their implications for the size evolution of both disks and spheroids,
in numerical hydrodynamic simulations (Choi et al. in prep) and
semi-analytic models (Somerville et al. in prep).

\subsubsection{Theoretical expectations for conditional size distributions}
\label{sec:sizedisp}

We found that the conditional distribution of P($\lambda R_h|m_*)$
from our SHAM is in remarkable agreement with the observed conditional
size distributions (size distribution in a bin of stellar mass and
redshift, $P(r_e|m_*)$). We found this to be the case in both the GAMA
and CANDELS samples in all redshift bins from $0.1<z<3$. This point
was already made in K13 with respect to nearby galaxies, but we have
shown it here more explicitly, in greater detail, and also for a
mass-selected high redshift galaxy sample.

This result is surprising for several reasons. First, in the
  context of semi-analytic models of disk size such as
  Eqn~\ref{eqn:mmw} above, we expect additional dispersion to arise
  from the terms depending on disk mass ($f_d$) and halo concentration
  ($c_{\rm NFW}$). Both of these quantities are expected to have
  significant halo-to-halo scatter. Indeed, \citet{desmond:2015} showed that a
  model based on abundance matching plus an angular momentum partition
  type model similar to Eqn~\ref{eqn:mmw} produces too large a scatter
  in galaxy size at fixed stellar mass.
Second, it holds across populations that are almost entirely disk
dominated to ones that are almost entirely composed of giant
ellipticals. This seems a non-trivial finding, given that disks are
rotation supported while spheroids are supported by velocity
dispersion.  It may be a coincidence, or it may tell us something
fundamental about the way that galaxies form. It may also appear
surprising in view of the large (roughly two orders of magnitude)
galaxy-to-galaxy scatter in the relationship between galaxy spin and
halo spin ($\lambda_{\rm galaxy}$ vs. $\lambda_{\rm h}$) predicted by
numerical simulations, as discussed in the introduction. However, we
emphasize that these two findings are not necessarily inconsistent,
although they do tell us something important about the physical
processes that shape galaxy structure.

The ratio $\lambda_{\rm galaxy}/\lambda_{\rm h}$ is equivalent to the
ratio of the specific angular momenta of the galaxy and the DM halo
$(J_{\rm gal}/M_{\rm gal})/(J_h/M_h) \equiv (j_{\rm gal}/j_{\rm h})$,
sometimes denoted $f_j$, and often adopted as a parameter in
semi-analytic models. For a disk, we can write (adopting the Bullock
definition of spin):
\begin{equation}
r_d = f_j \, \left(\frac{V_{\rm h}}{V_{\rm rot}}\right) (\lambda_{B} R_{\rm h})
\end{equation}
where $r_d$ is the exponential scale radius of the disk, $V_{\rm h}$
is the virial velocity of the halo, $V_{\rm rot}$ is the rotation
velocity of the disk, and other quantities are as defined
previously. It is clear from this example that an anti-correlation
between any of the terms (such as $f_j$ and $V_{\rm h}/V_{\rm rot}$ or
$f_j$ and $\lambda_h$, neither of which would be difficult to motivate
physically) could reconcile a large dispersion in $f_j$ with our
results.

It is also entirely possible that the \emph{distributions} of
$\lambda_{\rm galaxy}$ and $\lambda_{\rm h}$ could be similar, even if
their values are not well correlated for individual galaxies.  This
picture appears to be supported by the results of the
\citet{ceverino:2014} numerical hydrodynamic simulations (Dekel et
al. in prep). This could arise if, for example, the values of
$\lambda_{\rm galaxy}$ and $\lambda_{\rm halo}$ are determined by the
physical conditions at different times, or different spatial
locations.  In addition, \citet{burkert:2016} found that the
dispersion in galaxy spin parameter $\lambda_{\rm galaxy}$ for
observed star forming galaxies at redshift $\sim 0.8$--2 is similar to
the dispersion in halo spin parameters in dissipationless
simulations. Investigating whether comparable distributions of
$\lambda_{\rm galaxy}$ and $\lambda_{\rm halo}$ are indeed naturally
and generically produced in numerical cosmological simulations, and
better understanding the physical processes that lead to this result,
is an important issue to follow up.

Although we have included a simplified estimate of errors in the
stellar mass measurements in our SHAM, we have made no effort to
deconvolve the observational errors in the size measurements or to add
errors to the theoretical size predictions. The observational size
distributions are of course broadened by both size and stellar mass
measurement errors, implying that the theoretically predicted size
distributions are actually somewhat broader than the intrinsic
observed ones. However, it is also possible that the breadth of the
observational distributions is underestimated due to selection
effects. Galaxies with very large sizes may be missed due to surface
brightness selection effects (or their sizes underestimated), and
galaxies with very small sizes may be mistaken for stars, may be
unresolved, or may be preferentially discarded because the fit quality
is poor. One can see from Fig.~\ref{fig:sizedistev} that the most
compact galaxies predicted by the SHAM model are unresolved even by
WFC3 on HST in the higher redshift bins, or contain only a few pixels.

Our results highlight the importance of confronting the observed
conditional size distributions with predictions from state-of-the-art,
high resolution numerical hydrodynamic simulations, including a
detailed treatment of observational selection effects, errors, and
biases.

\section{Conclusions}
\label{sec:summary}

In this paper, we have explored an empirical approach for connecting
(in a statistical sense) the observed radii of the stellar bodies of
galaxies with the virial radii of their host dark matter halos. We
used a mapping between dark matter halo mass and stellar mass based on
sub-halo abundance matching (SHAM). We then explore observational
constraints on the mapping between galaxy effective radius and the
halo virial radius (SRHR). In addition, we explore the mapping between
galaxy radius and the product of the halo spin parameter and the halo
virial radius (SRHR$\lambda$). We find the following main results:

\begin{itemize}

\item At $z\sim 0.1$, the average ratio SRHR is consistent with being
  roughly independent of stellar mass, with a value of $\sim 0.018$
  over a broad range in stellar mass. Similarly, SRHR$\lambda$ is
  nearly independent of stellar mass with a value of $\sim 0.5$.

\item We find hints that SRHR and SRHR$\lambda$ have a stronger
  dependence on stellar mass at high redshift than locally, with high
  mass galaxies having a value of SRHR and SRHR$\lambda$ that is about
  50\% smaller than that of lower mass galaxies at $z\sim 2$. 

\item We find weak or negligible redshift evolution in SRHR over the
  interval $3<z<0.1$. For galaxies with stellar mass $m_* \lesssim
  10^{10.3} \msun$, SRHR has decreased by about a factor of 1.5 over
  this interval. For more massive galaxies, SRHR has \emph{increased}
  by a similar factor from $3<z<0.4$.

\item The preceding empirical results appear consistent with a picture
  in which massive galaxies at high redshift form via dissipative
  processes (such as gas-rich mergers or violent disk instabilities),
  leading to compact galaxies. As time progresses, galaxies become
  more gas poor, and massive galaxies seen closer to the present epoch
  are built up of a series of gas-poor mergers, leading to more
  extended stellar bodies.

\item The inferred redshift evolution of SRHR$\lambda$ depends on the
  definition of the spin parameter that is adopted.  If we adopt the
  Peebles definition of $\lambda$, we find a decrease of about a
  factor of 1.8 over $3<z<0.1$ for galaxies with $m_* \lesssim
  10^{10.3} \msun$, while if we use the Bullock definition of
  $\lambda$, we find results that are consistent with no significant
  change in SRHR$\lambda$ over this time period.

\item We find the conditional distribution of $\lambda R_h$ in stellar
  mass bins from our SHAM is in remarkably good agreement with the
  observed conditional size distributions in stellar mass bins from
  $z\sim 0.1$--3 (for both the GAMA and CANDELS samples). This
  suggests that there is little room for large galaxy-to-galaxy
  variations in SRHR$\lambda$, unless the width of the observed
  distribution is significantly underestimated, or internal
  correlations conspire to reduce the dispersion.

\item We caution that there are still significant uncertainties
  in key areas of this analysis, such as in converting from observed
  quantities (projected, light-weighted sizes) to intrinsic quantities
  (3D, stellar mass weighted sizes). These could impact our reported
  trends and further work is needed to constrain them better. 

\end{itemize}

Our results provide guidelines for adding size information to
SHAM-type models, as well as providing insights into the physics that
shapes galaxy sizes over cosmic time.

\section*{Acknowledgments}
We thank Seong-Kook Lee for helpful comments on the manuscript, and we
thank Aldo Rodr\'{i}guez-Puebla for useful discussions. We thank
Kuang-Han Huang for providing his data in electronic format and for
helpful discussions. We thank the anonymous referee for comments and
suggestions that improved the paper.  We acknowledge the Kavli
Institute for Theoretical Physics at the University of Santa Barbara,
where part of this work was performed. This research was supported in
part by the National Science Foundation under Grant No. NSF
PHY-1125915.  rss thanks the Downsbrough family for their generous
support, and acknowledges support from the Simons Foundation through a
Simons Investigator grant.  PB was supported by program number
HST-HF2-51353.001-A, provided by NASA through a Hubble Fellowship
grant from the Space Telescope Science Institute, which is operated by
the Association of Universities for Research in Astronomy,
Incorporated, under NASA contract NAS5-26555. PGP-G acknowledges
support from Spanish MINECO Grants AYA2015-63650-P and
AYA2015-70815-ERC.
We acknowledge the contributions of hundreds of
individuals to the planning and support of the CANDELS observations,
and to the development and installation of new instruments on HST,
without which this work would not have been possible. Support for HST
Programs GO-12060 and GO-12099 was provided by NASA through grants
from the Space Telescope Science Institute, which is operated by the
Association of Universities for Research in Astronomy, Inc., under
NASA contract NAS5-26555.

\bibliographystyle{mn}
\bibliography{mn-jour,rgrh}

\appendix
\section{Observed size-mass relations}
\label{sec:appendix:obs}

\begin{figure*} 
\begin{center}
\includegraphics[width=0.95\textwidth]{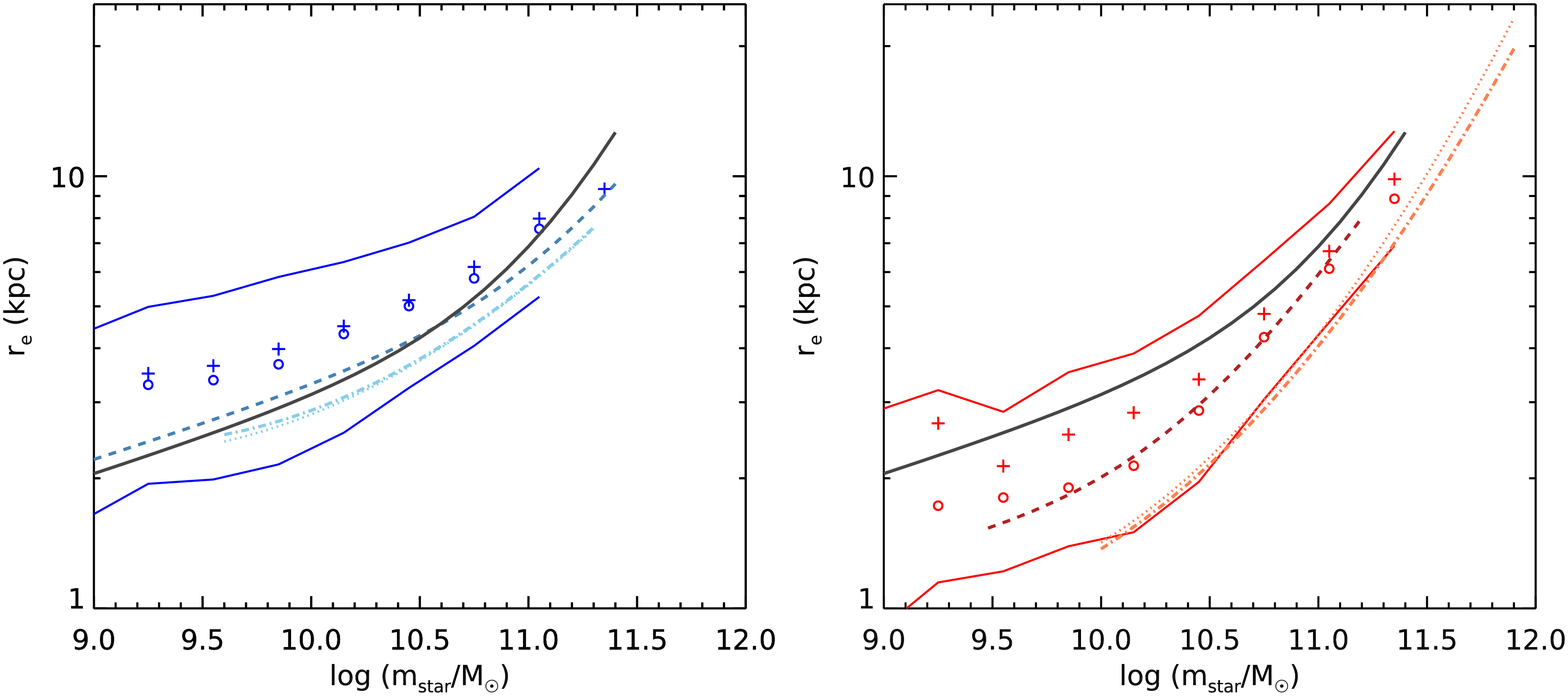}
\end{center}
\caption{Observed size-mass relation for the GAMA sample used in this
  analysis, compared with other relations from the literature. Here
  $r_e$ is the semi-major axis half-light radius in the $r$-band.
  Left panel: relation for disk-like galaxies with S{\'e}rsic
  parameter $n_s<2.5$. Crosses show means, while open circles show
  medians. Solid blue lines show the 16th and 84th percentiles. The
  gray-blue dashed line shows the fitting function for $n_s<2.5$
  galaxies given in \citet{lange:2015}, and light blue dotted and
  dot-dashed lines show the results for single S{\'e}rsic and
  S{\'e}rsic+exponential fits (respectively) from
  \citet{bernardi:2014}. Note that the \citet{bernardi:2014} sizes are
  circularized. Right panel: relation for spheroid-dominated galaxies
  with S{\'e}rsic parameter $n_s>2.5$. Lines and symbols are as in the
  left panel, but all for $n_s>2.5$ galaxies. The dark gray solid line
  in both panels shows the size-mass relation for the disk- and
  spheroid-dominated samples combined, converted to the 3D stellar
  half-mass radius using the type-dependent correction described in
  the text.  
  \label{fig:sizemass_gama}}
\end{figure*}

\begin{figure*} 
\begin{center}
\includegraphics[width=0.95\textwidth]{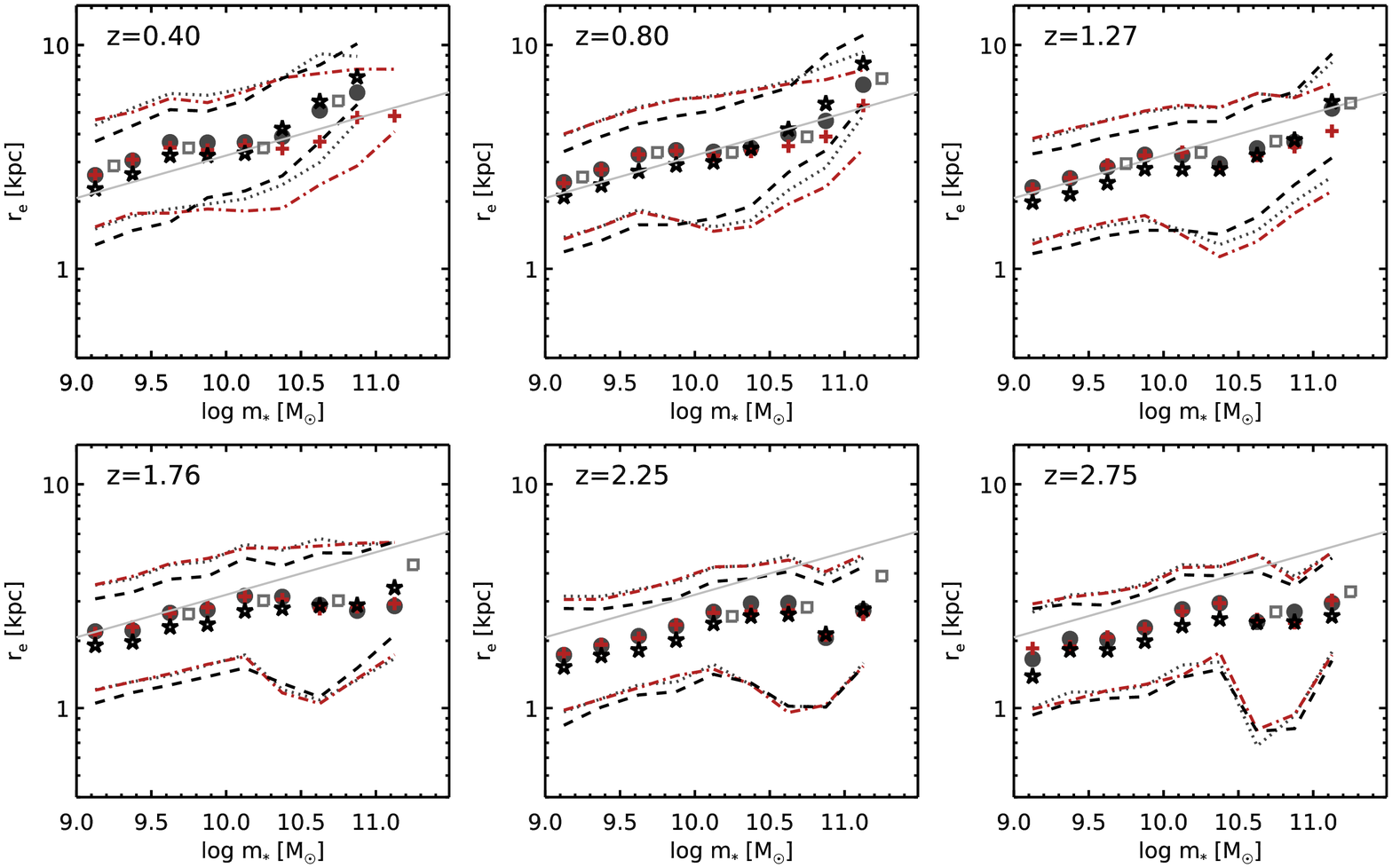}
\end{center}
\caption{Observed size-mass relation for the CANDELS sample used in
  this work, in redshift bins as indicated on the panels.  Dark-gray
  filled circles: results from our analysis. Open squares: results
  from the published analysis of vdW14. Red crosses: results from the
  analysis of H17. All of the preceding show median rest
  5000\AA\ half-light radii. The light gray line, repeated in each
  panel, shows the size mass relation from the $z=0.1$ GAMA sample.
  There is good agreement between the observed size-mass relation
  published by \protect\cite{vanderwel:2014}, that derived from our
  analysis of the CANDELS team catalogs, and that derived from the
  analysis of H17. Black stars: estimated median 3D stellar half-mass
  radii obtained by applying the type-dependent corrections described
  in the text to the CANDELS sample. }
\label{fig:sizemass_candels}
\end{figure*}

In Fig.~\ref{fig:sizemass_gama}, we show the size-mass relation for the
GAMA sample that we use in this work. We show the relation for
galaxies with S{\'e}rsic index $n_s<2.5$ (disk-dominated galaxies) and
$n_s>2.5$ (spheroid dominated galaxies) separately, as we apply
corrections for deprojection and to convert from half-light radii to
half stellar mass radii based on this division. We show a comparison
between the mean and median sizes in stellar mass bins from our
analysis and several relations from the literature, including the
analysis of GAMA by \citet{lange:2015} and the analysis of SDSS by
\citet{bernardi:2014}. The GAMA sizes appear to be systematically
larger at a given stellar mass than the SDSS sizes. Note that the
\citet{bernardi:2014} sizes are circularized, while the GAMA-based
sizes in our analysis and that of \citet{lange:2015} are semi-major
axis sizes. This could explain the offset for flattened galaxies, but
it is surprising that the offset appears similar for
spheroid-dominated galaxies, which should have nearly round isophotes.

In Fig.~\ref{fig:sizemass_candels} we show our derived size-mass
relations for the CANDELS sample used in this work. Here, we show the
full sample, without dividing into different galaxy types, but we
again apply a S{\'e}rsic-dependent correction for deprojection and to
convert from light to stellar mass. Our derived size-mass relation is
in excellent agreement with the published results of vdW14 and H17.

\section{Halo structural parameter definitions}
\label{sec:appendix:halos}

\begin{figure*} 
\begin{center}
\includegraphics[width=\textwidth]{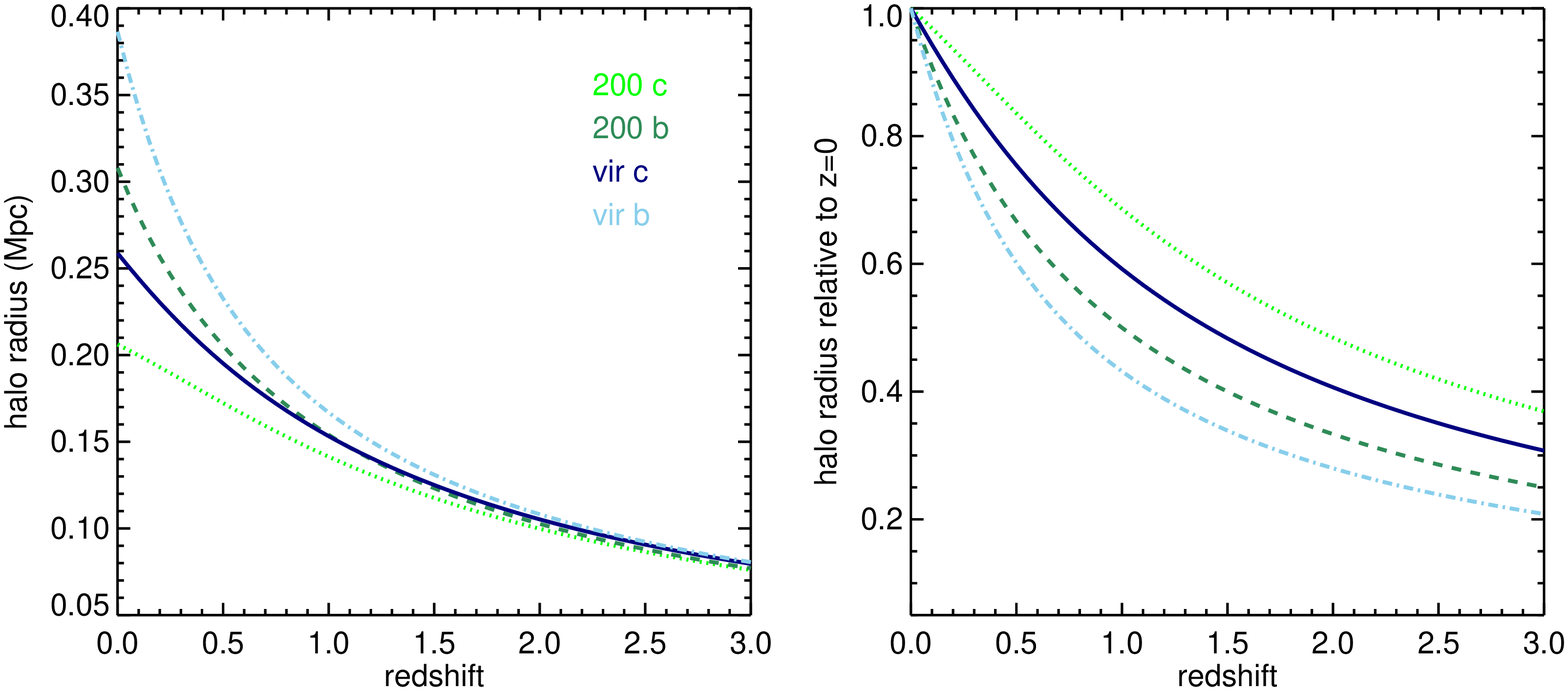}
\end{center}
\caption{Halo radius versus redshift for different definitions of halo
  mass and radius. Solid dark blue: $r_{\rm vir, crit}$; light blue
  dot-dashed: $r_{\rm vir, b}$; green dotted: $r_{\rm 200, crit}$;
  dark green dashed: $r_{\rm 200, b}$ (see text for definitions).
  Different definitions produce similar results at high redshift, but
  the halo radius may differ by as much as a factor of almost two in
  normalization at $z=0$ for different definitions, and the inferred
  evolution can differ by a similar amount.
\label{fig:rcomp}}
\end{figure*}

\begin{figure*} 
\begin{center}
\includegraphics[width=\textwidth]{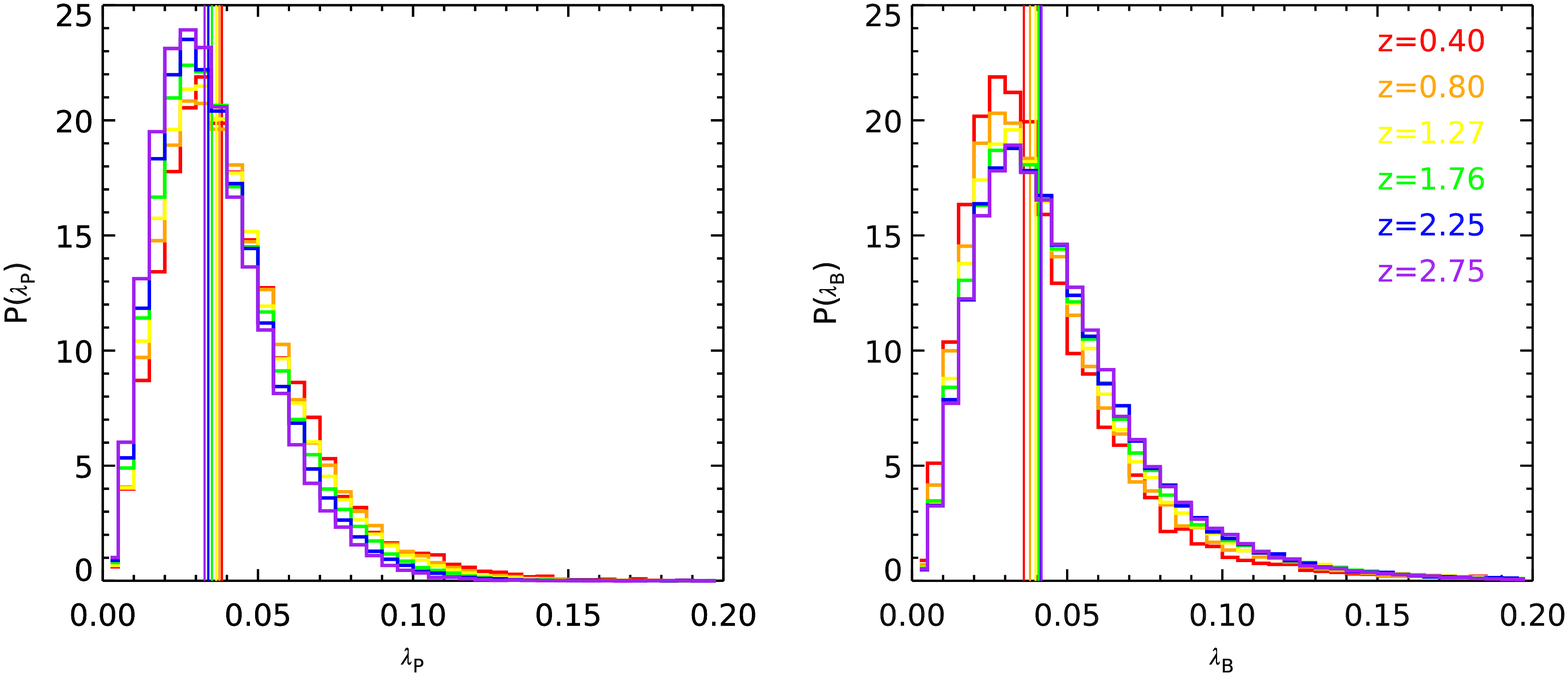}
\end{center}
\caption{Distributions of the spin parameter $\lambda$ in the
  Bolshoi-Planck simulations, in the redshift bins used in our
  analysis.  Left: Peebles definition $\lambda_P$. Right: Bullock
  definition $\lambda_B$. Vertical lines show the medians in each
  redshift bin. The median value of $\lambda_P$, along with the whole
  distribution, shifts to larger values with decreasing redshift,
  while the median value and distribution of $\lambda_B$ shift to
  smaller values with decreasing redshift.
\label{fig:spindist}} 
\end{figure*}

In this Appendix we show the differences between different definitions
of halo virial radius and spin parameter. It has become customary to
define dark matter halos as spherical overdensities within which the
average overdensity exceeds a threshold value. However, different
values of this overdensity parameter are used in the literature. The
most common conventions are to assume a fixed overdensity of 200 or to
assume a redshift dependent overdensity $\Delta_{\rm vir}$ as given in
\citet{bryan:1998}. To make matters even more confusing, some studies
apply the overdensity threshold relative to the critical density of
the Universe while others use the background density. This results in
different values of $R_h$ for a given $M_{h}$, different values of
halo number density (or abundance) at a given $M_h$, and different
redshift evolution for all quantities.  It also results in different
values for the total angular momentum of the halo, $J_h$, and spin
parameter $\lambda$.

In Fig.~\ref{fig:rcomp}, we show the virial radius as a function of
redshift for a halo with a mass of $10^{12}\msun$. We also show the
virial radius as a function of redshift at fixed mass, normalized to
the value at $z=0$. One can see from this figure that the halo radius
at a given mass differs at $z=0$ by as much as a factor of two in
different definitions, while all definitions produce nearly the same
value above $z\sim 3$. As a result, conclusions about the evolution of
halo radius across cosmic time can also differ by a similar
factor. The ``200 crit'' definition produces the least evolution,
while the ``vir background'' definition produces the most.

\begin{figure*} 
\begin{center}
\includegraphics[width=\textwidth]{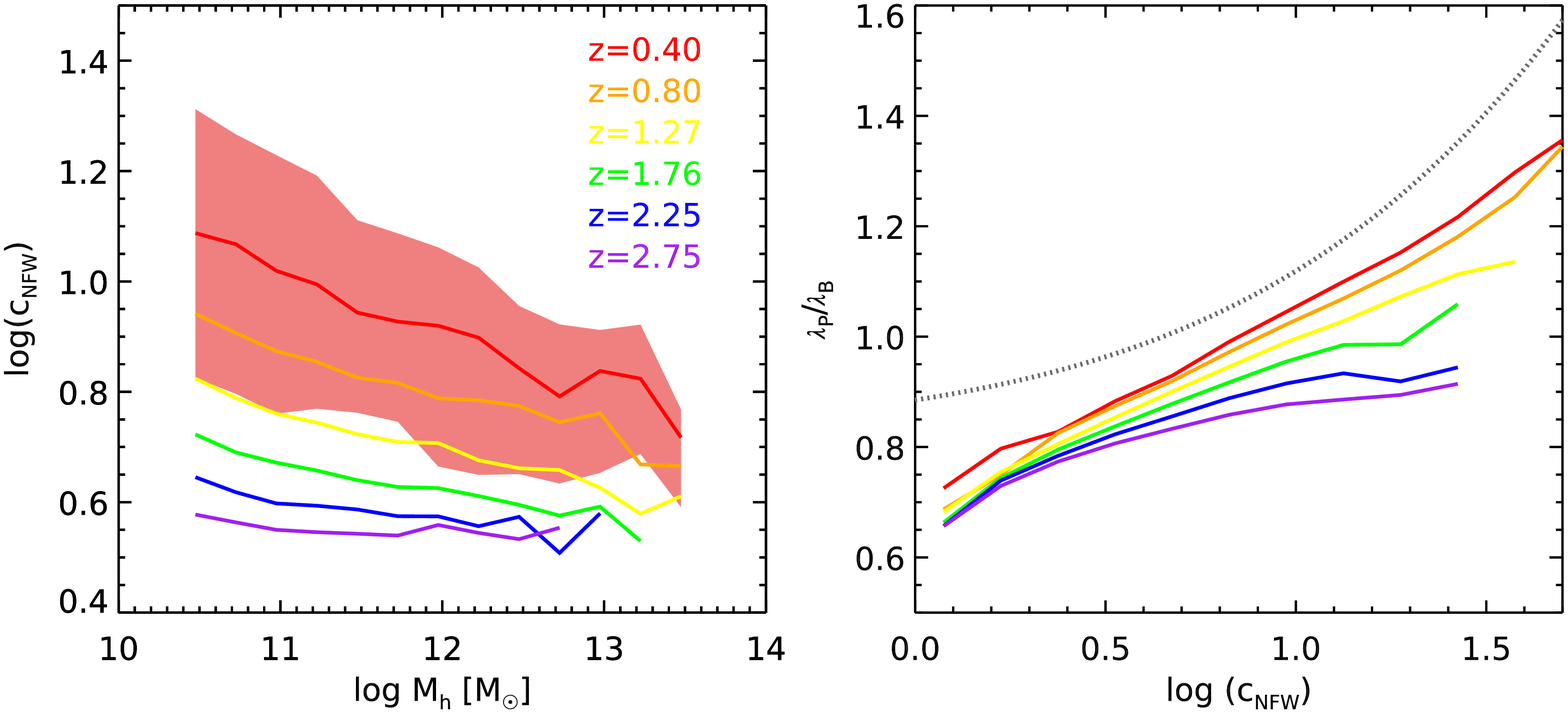}
\end{center}
\caption{Left: Median NFW concentration parameter versus halo mass for
  the same redshift bins used in our analysis. The shaded red area
  shows the 16 and 84th percentiles for the $z=0.1$--0.5 bin.  As is
  well known, halo concentrations at a fixed mass are lower at high
  redshift. Right: Ratio between the Peebles and Bullock definitions
  of the spin parameter in the Bolshoi-Planck halo catalogs as a
  function of the NFW concentration parameter. For perfect spherical
  halos with no sub-structure, with all particles on circular orbits,
  $\lambda_P/\lambda_B$ should be equal to the function $(f_c)^{1/2}$
  (see text). The evolution of halo concentrations go in the right
  direction to explain the differing evolution of $\lambda_P$ and
  $\lambda_B$ in the simulations, but do not appear to provide a
  complete explanation, as seen from the evolution in
  $\lambda_P/\lambda_B$ at fixed $c_{\rm NFW}$. 
\label{fig:haloprop}} 
\end{figure*}

The halo spin parameter clearly depends on the halo mass and radius
definition. In addition, two different dimensionless spin parameters
have been proposed in the literature, the ``Peebles'' and ``Bullock''
definitions described in \S\ref{sec:intro}. These are generally
assumed to be interchangable. In Fig.~\ref{fig:spindist} we show the
distributions of halo spin parameters in the Bolshoi-Planck
simulations, in the six redshift bins used in the rest of our analysis
($0.1$--$0.5$, $0.5$--$1.0$, $1.0$--$1.5$, $1.5$--$2.0$, $2.0$--$2.5$,
and $2.5$--$3.0$; the redshift labels on the plots indicate the volume
mid-point of each bin). We show the distributions for both the Peebles
and Bullock definition of the halo spin parameter. In both
Fig.~\ref{fig:spindist} and Fig.~\ref{fig:haloprop}, we show only
``distinct'' halos (halos that are not a sub-halo of another halo)
with $M_h > 10^{10.35}\, \msun$ (this is the mass limit quoted by
\citet{rodriguez-puebla:2016} for robust determination of halo
structural properties in the Bolshoi-Planck simulations).

We see that the distribution of $\lambda_P$ shifts towards larger
values at lower redshift, while the distribution of $\lambda_B$ shifts
towards smaller values at lower redshift. The ratio of the median
value of $\lambda_P$ in the $2.5<z<3.0$ bin to that in the $0.1<z<0.5$
bin is 0.85, while this ratio is 1.16 for $\lambda_B$.

The original motivation behind the Peebles definition of the spin
parameter $\lambda_P$ was to represent the fraction of the total
energy of a system in the form of ordered rotational motion. Thus the
total energy $E$ comes in to the expression. In the special case of a
truncated singular isothermal sphere in which all particles are on
circular orbits,
\begin{equation}
  E = -\frac{GM^2}{2R} = -\frac{MV_c^2}{2}
\end{equation}
Thus we see that for truncated singular isothermal spheres with all
particles on circular orbits, the Peebles and Bullock definitions of
spin are the same ($\lambda_P = \lambda_B$). For the more
cosmologically relevant (but still simplified) case of perfect NFW
spheres with all particles on circular orbits, the total energy $E$ is
\begin{equation}
  E = -\frac{GM^2}{2R} f_c
\end{equation}
where the function $f_c$ depends only on the NFW concentration
parameter (see e.g. Eqn. 23 of \citet{mo:1998}). This function can be
well-approximated by the fitting function \citep{mo:1998}
\begin{equation}
  f_c \simeq \frac{2}{3} + \left(\frac{c_{\rm NFW}}{21.5}\right)^{0.7} .
  \label{fc}
\end{equation}
Thus, for smooth, spherical NFW halos with all particles on circular
orbits, $\lambda_P/\lambda_B = (f_c)^{1/2}$.

Fig.~\ref{fig:haloprop} shows the median halo mass versus NFW
concentration parameter for halos in the Bolshoi-Planck catalogs in
the same redshift bins. As is well known, the concentration for a
fixed halo mass is lower at high redshift. As $f_c$ is a monotonically
increasing function of the concentration parameter $c_{\rm NFW}$, we
would therefore expect $\lambda_P/\lambda_B$ to be lower at high
redshift as well (for fixed halo mass). However, we see from the right
panel of Fig.~\ref{fig:haloprop} that the explanation appears to be
somewhat more complicated: halos have lower median values of
$\lambda_P/\lambda_B$ at high redshift even at fixed
concentration. This could be due to halos in cosmological simulations
deviating from sphericity, having sub-structure, or deviating from
having all particles on circular orbits, by different amounts at
different redshifts. Clearly, this is an interesting issue to
investigate in more detail, as it may have important implications for
the structural properties of galaxies and their evolution.

\end{document}